\documentclass[twocolumn,twocolappendix]{aastex63}
\usepackage{commath} 
\usepackage{gensymb} 
\usepackage{eucal} 
\usepackage{aas_macros}
\usepackage[caption=false]{subfig} 
\usepackage{float}
\usepackage{enumerate}
\usepackage{natbib}
\usepackage{hyperref}

\usepackage{newtxtext,newtxmath}

\usepackage[T1]{fontenc}
\usepackage{ae,aecompl}
\usepackage{comment}

\usepackage{graphicx,tabularx}	
\usepackage{amsmath}	
\usepackage{amssymb}	
\usepackage{booktabs}
\usepackage[inline,shortlabels]{enumitem}
\setlist{itemjoin* = { and\enspace}}
\usepackage{color}
\usepackage{multirow}

\received{2022 August 31}
\revised{2023 January 18}
\accepted{2023 January 31}
\submitjournal{ApJ}

\renewcommand{\vec}[1]{\boldsymbol{#1}}

\newcommand{\mat}[1]{\boldsymbol{#1}}

\newcommand{\m}{\ensuremath{\,{\rm m}}}

\newcommand{\K}{\ensuremath{\, {\rm K}}}

\shorttitle{Semi-blind Foreground Subtraction}
\shortauthors{Zuo, Chen \& Mao}

\graphicspath{{./}}

\begin{document}


\title{A Semi-blind PCA-based Foreground Subtraction Method for 21~cm Intensity Mapping}

\correspondingauthor{Shifan Zuo, Xuelei Chen, Yi Mao}
\email{sfzuo@bao.ac.cn (SZ), xuelei@cosmology.bao.ac.cn (XC), ymao@tsinghua.edu.cn (YM)}

\author[0000-0003-3858-6361]{Shifan Zuo}
\affiliation{Department of Astronomy, Tsinghua University, Beijing 100084, China}
\affiliation{Key Laboratory of Computational Astrophysics, National Astronomical Observatories, Chinese Academy of Sciences, Beijing 100101, China}

\author[0000-0001-6475-8863]{Xuelei Chen}
\affiliation{Key Laboratory of Computational Astrophysics, National Astronomical Observatories, Chinese Academy of Sciences, Beijing 100101, China}
\affiliation{School of Astronomy and Space Science, University of Chinese Academy of Sciences, Beijing 100049, China}
\affiliation{Department of Physics, College of Sciences, Northeastern University, Shenyang 110819, China}
\affiliation{Center of High Energy Physics, Peking University, Beijing 100871, China}

\author[0000-0002-1301-3893]{Yi Mao}
\affiliation{Department of Astronomy, Tsinghua University, Beijing 100084, China}


\begin{abstract}
The Principal Component Analysis (PCA) method and the Singular Value Decomposition (SVD) method are widely used for foreground subtraction in 21~cm intensity mapping experiments. We show their equivalence, and point out that the condition for completely clean separation of foregrounds and cosmic 21~cm signal using the PCA/SVD is unrealistic.  
We propose a PCA-based foreground subtraction method, dubbed ``Singular Vector Projection (SVP)'' method, which exploits {\it a priori} information of the left and/or right singular vectors of the foregrounds. We demonstrate with simulation tests that this new, semi-blind method can reduce the error of the recovered 21~cm signal by orders of magnitude, even if only the left and/or right singular vectors in the largest few modes are exploited. The SVP estimators provide a new, effective approach for 21~cm observations to remove foregrounds and uncover the physics in the cosmic 21 cm signal. 


\end{abstract}

\keywords{Astronomical methods (1043); Cosmology (343); H I line emission (690); Principal component analysis (1944)}


\section{Introduction}\label{S:intro}
The 21~cm radiation due to the hyperfine transition of atomic hydrogen \citep{Furlanetto2006,Morales2010,Pritchard2012} has emerged as a promising cosmological probe that can be used to reconstruct the history from cosmic dawn to the epoch of reionization, as well as the large-scale structure of the Universe after reionization. 
A number of completed and ongoing radio interferometric array experiments are targeting the 21~cm intensity mapping and/or its power spectrum measurement, including the Precision Array for Probing the Epoch of Reionization (PAPER; \citealp{Parsons2010}), the Giant Meterwave Radio Telescope (GMRT; \citealt{Paciga2013,2017A&A...598A..78I}), the Murchison Widefield Array (MWA; \citealt{Bowman2013}), the LOw Frequency Array (LOFAR; \citealt{Patil2017,Gehlot2019}), the Hydrogen Epoch of Reionization Array (HERA; \citealt{DeBoer2017}), the Green Bank Telescope (GBT; \citealt{Chang2010,Masui2013,Switzer2013}), the Parkes Radio Telescope \citep{Anderson2018}, the Owens Valley Long Wavelength Array (OVRO-LWA; \citealt{Eastwood2018,Eastwood2019}), the MeerKAT telescope \citep{2021MNRAS.505.3698W}, the Canadian Hydrogen Intensity Mapping Experiment (CHIME; \citealt{Newburgh2014,2022arXiv220201242C}), the Hydrogen Intensity and Real-time Analysis eXperiment (HIRAX; \citealt{2022JATIS...8a1019C}), the Tianlai experiment \citep{Chen2012,Xu2015,Li:2020ast,Wu:2020jwm}, and the BAO from Integrated Neutral Gas Observations (BINGO; \citealt{Battye2012}). 
Next generation experiments with higher sensitivities are also under construction, including the Canadian Hydrogen Observatory and Radio-transient Detector (CHORD; \citealt{Vanderlinde2019}), and the Square Kilometre
Array (SKA; \citealt{Koopmans2015,Maartens2015}). 

Detection of the cosmic 21~cm signal is technically very challenging, even for next-generation interferometric arrays with large receiving area, because the astrophysical foregrounds from the galactic and extragalactic sources are four to five orders of magnitude larger than the cosmic 21~cm signal.  Advanced techniques for foreground subtraction or avoidance have been developed (see, e.g. the review of \citealt{2020PASP..132f2001L}). However, residual foreground still remains a main source of systematics for 21~cm intensity mapping. These methods include the low order polynomial fitting \citep{Wang2006,Jelic2008,Liu2009a,Liu2009b}, the Principal Component Analysis (PCA) method \citep{Masui2013,Alonso2015,Sazy2015}, the Singular Value Decomposition (SVD) method \citep{Switzer2013,Switzer2015}, the robust PCA method \citep{Zuo2019}, the Independent Component Analysis (ICA) method
\citep{Chapman2012,Wolz2014,Alonso2015}, the Generalized Morphological Component Analysis (GMCA) method \citep{Chapman2013}, and the Gaussian Process Regression (GPR) method \citep{Mertens2018}. 

In particular, the PCA method has been shown to work well in simulation tests \citep{Alonso2015,Sazy2015} and is applied extensively to the analysis of observational data \citep{Masui2013,2022arXiv220601579C}. It has been demonstrated using the mock data from simulations that the residual foreground can be effectively suppressed to the level comparable to, or even below, that of cosmic 21~cm signal, after removing four to five PCA modes (see, e.g. \citealt{Alonso2015,Zuo2019}). Those demonstrations often only assume that the spectrum of foreground is smooth in the frequency direction. Furthermore, it is shown that satisfactory performance can be obtained by removing six to seven PCA modes, if we take into account complex instrumental effects such as frequency-dependent beam, $1/f$ noise and other imperfections \citep{Alonso2015,Sazy2015}. \citet{Masui2013} made a measurement of the 21~cm brightness fluctuation power spectrum at $z\sim 0.8$ by the cross-correlation of 21~cm signal using GBT with the large-scale structure traced by galaxies that are optically detected in the WiggleZ Dark Energy Survey. Nevertheless, it is necessary to subtract out 20 PCA modes in the GBT observations in order to suppress the
residual foreground to an acceptably low level \citep{Masui2013}. Moreover, in a recent, similar detection of the cross-correlation of 21~cm signal using MeerKAT with galaxies from the WiggleZ data, 30 PCA modes are removed \citep{2022arXiv220601579C}, due to the modulation of instrumental responses such as calibration imperfections, residual Radio Frequency Interference (RFI) and other systematics. 

Removing such a large number of PCA modes results in severe signal loss and foreground mixing. In fact, even after removal of 30 PCA modes, the resulting 21~cm signal in MeerKAT has not reached the estimated auto-power spectrum level \citep{2022arXiv220601579C}, which implies that there are still residual RFI, foregrounds, and other systematics in the 21~cm data. To avoid these additive biases, \citet{Masui2013,2022arXiv220601579C} employed the cross-correlation technique of the 21~cm signal with galaxy surveys for the detection of the 21~cm brightness fluctuations. This difficulty suggests that the PCA\footnote{The GPR method has recently been employed widely as a new foreground removal technique, e.g., by LOFAR \citep{Gehlot2019,Mertens2020} and HERA \citep{Ghosh2020}. It is shown in the context of low-redshift H{~\sc i} intensity mapping with single-dish experiments that GPR performs better on small scales in recovering the H{~\sc i} power spectrum than PCA. However, it is still not at a position to claim that the PCA is obsolete, because the recovery results of GPR are not so good as those of PCA in some cases, e.g.~when there are missing frequency channels due to RFI flagging \citep{Soares2022}. } 
method needs to be further improved in order to remove the foregrounds to such a cleaned level that we can realize the detection of the cosmic 21~cm signal through the auto-power spectrum. 

The standard PCA method is a {\it blind} method, in the sense that no prior information about the signal or foreground is assumed other than the assumption of smooth spectrum of the foregrounds in frequency, which is too generic. Some {\it semi-blind} methods have been proposed to make assumptions about the foregrounds and/or signal that are not so strong as assuming an ansatz of foregrounds in the polynomial fitting method, and demonstrated to alleviate the problem of signal loss in foreground subtraction. These methods include the 
Karhunen-Lo\`eve (KL) transform method \citep{Shaw2014,Shaw2015} that employs the modeled foregrounds and signal covariance matrices to form a set of modes that are ordered by their signal-to-contaminant ratios, and the Generalized Needlet Internal Linear Combination (GNILC) method \citep{Olivari2016} that uses the H{~\sc i} covariance matrix to separate the foregrounds from the signal in a wavelet (or needlet) space. 
While it is indeed easier to model the covariance matrix of the foregrounds and signal than the foregrounds or the signal {\it per se}, the overall magnitude of the covariance matrix may be highly biased, because the foregrounds are four to five orders of magnitude stronger than cosmic 21~cm signal. If we take into account the thermal noise, modeling the covariance matrix can be even more challenging due to the correlated noise between different frequencies and/or between different pixels induced by instrumental effects. These issues impose practical difficulty in applying these methods to the observational data. 

In this paper, instead, we will propose a new semi-blind PCA-based method for foreground subtraction, dubbed {\it Singular Vector Projection} (SVP). The SVP method assumes that the covariance matrix of foregrounds can be modeled {\it a priori} by theoretical modeling or from observational data, and then the left and/or right singular vectors of the foregrounds can be obtained by eigen-decomposition of the covariance matrix of the foregrounds in frequency and/or pixel space. Note that the overall magnitude of the covariance matrix affects its eigenvalues, but not the left and right singular vectors. If we can design the SVP estimators in such a manner that they contain the singular vectors, but not the eigenvalues, as will be shown in the paper below, then the estimators are independent of the overall magnitude of the covariance matrix. In other words, the estimators would not be affected if the overall magnitude of the covariance matrix was biased. We will also demonstrate the effectiveness of SVP in reducing the signal loss and foreground mixing in foreground subtraction.  

The rest of this paper is organized as follows. In \S\ref{S:mo}, we review the PCA and SVD methods for foreground subtraction briefly, show their equivalence, and derive the ideal condition for completely clean separation of foregrounds and cosmic 21~cm signal (i.e.\ without any signal loss or foreground mixing). We introduce our SVP method for foreground subtraction in \S\ref{S:sigvec}, and demonstrate its effectiveness with simulation tests in \S\ref{S:ex}. We make concluding remarks in \S\ref{S:con}. We leave some technical details (the mathematical proof of the inequalities for signal losses) to Appendix~\ref{S:csl}. 

\section{The PCA and SVD methods} 
\label{S:mo}

In this section, we briefly review the standard PCA and SVD methods for foreground subtraction. We will show their equivalence, and prove the ideal, yet unrealistic, conditions for completely clean separation of foregrounds and cosmic 21~cm signal. 

\subsection{Problem Setup} 

The 21~cm observational data is presented as a 3D image cube with
two angular directions and one frequency direction. The two angular
directions can be combined into a single dimension of size $p$, where $p$ is the
number of image pixels, and therefore the dataset is represented  in form of a matrix $\mat{D} \in \mathbb{R}^{n \times  p}$, where $n$ is the number of frequency bins. This general representation of the dataset is valid for the observation of a patch of sky as well as the full sky. Without losing generality, we assume $n \le p$ in the analysis below, i.e.\ the number of frequency bins is equal to or less than the number of pixels in the map. This is often the case, but the conclusions herein are also valid for the case of $n > p$.

The observational data is a linear combination of the foregrounds, cosmic 21~cm signal, and systematic noise, by writing it as 
\begin{equation} \label{eq:DFN}
  \mat{D} = \mat{F} + \mat{N},
\end{equation}
where the total foregrounds are denoted as $\mat{F}$, and $\mat{N}$ includes both cosmic 21~cm signal and noise.

\subsection{The PCA Method} \label{S:pca}

We follow \citet{Alonso2015,Sazy2015,Masui2013} for the review of the PCA method for foreground subtraction. Consider the covariance matrix $\mat{R}\in \mathbb{R}^{n \times  n}$ of the dataset in frequency space, $\mat{R} = \mat{D} \mat{D}^{\rm T} $, and perform the eigen-decomposition 
\begin{equation} \label{eq:RDD}
  \mat{R} = \mat{U} \mat{\Lambda} \mat{U}^{\rm T}\,.
\end{equation}
Here, $\mathcal{O}^{\rm T}$ denotes the transpose of a matrix $\mathcal{O}$. $\mat{\Lambda}$ is a $n \times n$ diagonal matrix in which the diagonal elements are the eigenvalues $\left\{\lambda_{i}\right\}$ of the matrix $\mat{R}$. The matrix $\mat{U}$ is a $n \times n$ real orthogonal matrix in which its $i^{\rm th}$ column is the eigenvector of $\mat{R}$ corresponding to the $i^{\rm th}$ eigenvalue $\lambda_{i}$.
The magnitude of $\lambda_{i}$ gives the variance of the corresponding eigen-mode, and 
each eigenvalue measures the contribution of its
corresponding eigenvector to the total sky variance.  
Since the foregrounds dominate the full data overwhelmingly, we can
project out the dominant components by picking up the $m$ largest
eigenvalues and their corresponding eigenvectors, so the foregrounds and the 21~cm signal can be estimated, respectively, by 
\begin{eqnarray}
  \mat{F}_{\rm PCA} &=& \mat{U} \mat{\Pi}_{n,m} \mat{U}^{\rm T} \mat{D}\,, \label{eq:Fp} \\
  \mat{N}_{\rm PCA} &=& \mat{D} - \mat{F}_{\rm PCA} = \mat{U} (\mat{I}_{n} - \mat{\Pi}_{n,m}) \mat{U}^{\rm T} \mat{D}\,. \label{eq:Np}
\end{eqnarray} 
Here, $\mat{I}_{n}$ is a $n \times n$ identity matrix, and $\mat{\Pi}_{n,m}$ is a projection matrix from dimension $n$ to $m$, i.e.\ a $n \times n$ diagonal matrix in which $m$ diagonal elements are unity if they correspond to the picked eigenvalues, and all other diagonal elements are zero. 

\subsection{The SVD Method} \label{S:svd}

The dataset $\mat{D}$ can be decomposed with SVD as 
\begin{equation} \label{eq:Dsvd}
  \mat{D} = \mat{U}' \mat{S} \mat{V}^{\rm T}\,.
\end{equation}
Here, $\mat{S} \in \mathbb{R}^{k \times k}$ is a diagonal matrix in which its positive diagonal elements $\left\{s_{i}\right\}$ are the singular values, where the integer $k \le
\text{min}(n, p)$ is the number of singular values of the dataset. The matrices 
$\mat{U}' \in \mathbb{R}^{n \times k}$ and $\mat{V} \in \mathbb{R}^{p \times k}$ are the
corresponding left and right singular vectors, respectively. These two singular-vector matrices satisfy the conditions, $\mat{U}'^{\rm T}
\mat{U}' = \mat{I}_{k}$ and $\mat{V}^{\rm T} \mat{V} = \mat{I}_{k}$,
where $\mat{I}_{k}$ is a $k \times k$ identity matrix. However, in general, they do not necessarily meet the following conditions, $\mat{U}' \mat{U}'^{\rm T} = \mat{I}_{n}$ (unless $k = n$), and $\mat{V} \mat{V}^{\rm T} = \mat{I}_{p}$ (unless $k = p$). For this reason, 
a matrix like $\mat{U}'$ and $\mat{V}$ is called a {\it partial} orthogonal matrix, because it contains some columns of an orthogonal matrix. 

Using SVD, the foregrounds can be projected out by picking up the largest $m$ singular value modes \citep{Switzer2013,Switzer2015}, similar to the PCA. The foregrounds and the 21~cm signal can be estimated, respectively, by 
\begin{eqnarray} 
  \mat{F}_{\text{SVD}} &=& \mat{U}' \mat{\Pi}_{k,m} \mat{S} \mat{V}^{\rm T}\,, \label{eq:Fsvd} \\
  \mat{N}_{\text{SVD}} &=& \mat{D} - \mat{F}_{\text{SVD}} = \mat{U}' (\mat{I}_{k} - \mat{\Pi}_{k,m}) \mat{S} \mat{V}^{\rm T}\,. \label{eq:Nsvd}
\end{eqnarray}

\subsection{The Equivalence of PCA and SVD}

Note that the dimension $k$ of the matrix $\mat{S}$ is also the number of positive eigenvalues for the covariance matrix $\mat{R}$. The eigenvalues of $\mat{R}$ are non-negative, which means that there are $n-k$ zero eigenvalues of $\mat{R}$. Below, we assume that the $k$ positive eigenvalues of $\mat{R}$ (or equivalently $k$ singular values of $\mat{S}$) are in descending order. 

Substituting Eq.~(\ref{eq:Dsvd}) to Eq.~(\ref{eq:RDD}), we get $\mat{U} \mat{\Lambda} \mat{U}^{\rm T} = \mat{U}' \mat{S}^{2} \mat{U}'^{\rm T}$. 
It is straightforward to prove that $\mat{U} \mat{\Lambda} \mat{U}^{\rm T} = \mat{U}_{k} \mat{\Lambda}_{k} \mat{U}^{\rm T}_{k}$, where $\mat{\Lambda}_{k}$ is the $k \times k$ subset of $\mat{\Lambda}$, i.e.\ the diagonal matrix in which the diagonal elements are the $k$ positive eigenvalues of $\mat{R}$, and $\mat{U}_{k}$ is the $n \times k$ subset of $\mat{U}$, i.e.\ the matrix in which its columns are the first $k$ columns of $\mat{U}$ corresponding to the $k$ positive eigenvalues. So we find  $\mat{\Lambda}_{k} = \mat{S}^{2}$ and $\mat{U}_{k} = \pm \mat{U}'$. The undetermined sign is due to the fact $\mat{U}' \mat{S} \mat{V}^{\rm T} = (-\mat{U}') \mat{S} (-\mat{V}^{\rm T})$. 
Ignoring this sign degeneracy, we have 
\begin{equation}
    \mat{U}' = \mat{U}_{k}\,.  \label{eq:id0}
\end{equation}


 
It is straightforward to prove that 
\begin{eqnarray}
\mat{U} \mat{\Pi}_{n,m} \mat{U}^{\rm T} &=& \mat{U}_{k}  \mat{\Pi}_{k,m} \mat{U}^{\rm T}_{k}\,, \label{eq:id1}\\
\mat{D} = \mat{U} \mat{U}^{\rm T} \mat{D} &=& \mat{U}_{k} \mat{U}_{k}^{\rm T} \mat{D} \,.\label{eq:id2}
\end{eqnarray}
The proof uses the identity $\mat{U}_{k}^{\rm T} \mat{U}_{k} = \mat{I}_{k}$. Substituting Eq.~(\ref{eq:Dsvd}) to Eqs.~(\ref{eq:Fp}) and \ref{eq:Np}, therefore, we find $ \mat{F}_{\text{PCA}} = \mat{F}_{\text{SVD}} $, and $ \mat{N}_{\text{PCA}}= \mat{N}_{\text{SVD}}$. This shows that foreground subtraction results using PCA and SVD methods are equivalent. 

As such, the estimators in the PCA/SVD are 
\begin{eqnarray}
  \mat{F}_{\rm PCA/SVD} &=& \mat{U} \mat{\Pi} \mat{U}^{\rm T} \mat{D} = \mat{U} \mat{\Pi} \mat{S} \mat{V}^{\rm T}\,, \label{eq:Fpv} \\
  \mat{N}_{\rm PCA/SVD} &=& \mat{U} (\mat{I} - \mat{\Pi}) \mat{U}^{\rm T} \mat{D} = \mat{U} (\mat{I} - \mat{\Pi}) \mat{S} \mat{V}^{\rm T}\,. \label{eq:Npv}
\end{eqnarray} 
For simplicity, hereafter in this paper, we will drop the prime ($'$) in $\mat{U}'$ (because of Eq.~\ref{eq:id0}), and drop the subscripts in $\mat{I}$, $\mat{\Pi}$, and $\mat{U}$, because the dimensionalities of these matrices can be understood both in the context of PCA (where $\mat{U}$ is a $n \times n$ matrix) and in the context of SVD (where $\mat{U}$ is a $n \times k$ matrix) due to Eqs.~(\ref{eq:id1}) and (\ref{eq:id2}), as long as the interpretation of dimensionalities are self-consistent.




\subsection{Conditions for Completely Cleaned Separation} \label{S:cond}

In this subsection, we attempt to answer this question in the framework of PCA/SVD method: under what condition can we separate the foregrounds from the 21~cm signal in such a completely cleaned manner that there is no signal loss or foreground mixing? 

Consider the SVD of the foregrounds and the signal, respectively, 
\begin{eqnarray}
    \mat{F} & = & \mat{U}_{f} \mat{S}_{f} \mat{V}^{\rm T}_{f}\,\\
    \mat{N} &=&  \mat{U}_{n} \mat{S}_{n} \mat{V}^{\rm T}_{n}\,.
\end{eqnarray}
The subscripts $f$ and $n$ denote the foreground and signal (with noise), respectively.  

If the foregrounds and the signal are fully separated, then 
\begin{align}
 \mat{U}_{f} \mat{S}_{f} \mat{V}^{\rm T}_{f} &= \mat{U} \mat{\Pi}  \mat{S} \mat{V}^{\rm T}, \notag \\
 \mat{U}_{n} \mat{S}_{n} \mat{V}^{\rm T}_{n} &= \mat{U} (\mat{I} - \mat{\Pi}) \mat{S} \mat{V}^{\rm T}. \notag
\end{align}

Since the SVD is unique, the completely cleaned separation can be realized if and only if the singular values of $\mat{F}$ are the largest $m$ positive singular values of $\mat{D}$, and the columns of $\mat{U}_{f}$ ($\mat{V}_{f}$) are the corresponding singular
vectors in $\mat{U}$ ($\mat{V}$), 
while the singular values of $\mat{N}$ are the remaining positive singular values
of $\mat{D}$, and the columns of $\mat{U}_{n}$ ($\mat{V}_{n}$) are the corresponding singular vectors in $\mat{U}$ ($\mat{V}$). 

Thus, we find the conditions: (1) $\mat{U}_{f}^{\rm T} \mat{U}_{n} = \mat{0}$; (2) $\mat{V}_{f}^{\rm T} \mat{V}_{n} = \mat{0}$; (3) $\min{\mat{S}_{f}} > \max{\mat{S}_{n}}$. 
The first two conditions are equivalent to the orthogonality conditions
\begin{align} \label{eq:FN}
  \mat{F}^{\rm T} \mat{N} &= \mat{N}^{\rm T} \mat{F} = \mat{0}, \notag \\
  \mat{F} \mat{N}^{\rm T} &= \mat{N} \mat{F}^{\rm T} = \mat{0}.
\end{align}
The first condition means that there is no pixel-wise cross-correlation between the foregrounds and the signal, so the pixel covariance can completely separate the contributions from the foregrounds and from the signal, i.e.\ $\mat{D}^{\rm T} \mat{D} = \mat{F}^{\rm T} \mat{F} + \mat{F}^{\rm T} \mat{N} + \mat{N}^{\rm T} \mat{F} + \mat{N}^{\rm T} \mat{N} =  \mat{F}^{\rm T} \mat{F} + \mat{N}^{\rm T} \mat{N} $. Similarly, the second condition means that there is no frequency-wise cross-correlation between the foregrounds and the signal.

Eq.~(\ref{eq:FN}) is the necessary condition for complete separation, but there is another implicit condition in practice --- the number $m$ of PCA/SVD modes of the foregrounds should be known {\it a priori}. If $m$ is known, the foregrounds can be reconstructed from the left and right singular vectors corresponding to the largest $m$ singular values of $\mat{D}$, and the signal is recovered from the remaining singular vectors. 

However, we note that these conditions for complete separation of foregrounds and signal are ideal and unsatisfied in most cases. As the result, the signal loss and foreground mixing are unavoidable in practice.

\subsection{Signal Loss, Foreground Mixing and Recovery Error}


Substituting Eq.~(\ref{eq:DFN}) to Eq.~(\ref{eq:Npv}), the estimated signal with the PCA/SVD is
\begin{align} \label{eq:Np2}
  \mat{N}_{\rm PCA/SVD} &= \mat{U} (\mat{I} - \mat{\Pi}) \mat{U}^{\rm T} \mat{F} +
  \mat{U} (\mat{I} - \mat{\Pi}) \mat{U}^{\rm T} \mat{N} \notag \\
           &= \mat{U} (\mat{I} - \mat{\Pi}) \mat{U}^{\rm T} \mat{U}_{f}
           \mat{S}_{f} \mat{V}_{f}^{\rm T} + \mat{U} (\mat{I} -
           \mat{\Pi}) \mat{U}^{\rm T} \mat{U}_{n} \mat{S}_{n} \mat{V}_{n}^{\rm T}.
\end{align}
For PCA/SVD, the {\it foreground mixing} is
\begin{equation} \label{eq:Fmix}
  \mat{F}^{\text{mix}}_{\rm PCA/SVD} = \mat{U} (\mat{I} - \mat{\Pi}) \mat{U}^{\rm T} \mat{F},
\end{equation}
and the {\it signal loss} is
\begin{equation} \label{eq:Nloss}
  \mat{N}^{\text{loss}}_{\rm PCA/SVD} = \mat{N} - \mat{U} (\mat{I} - \mat{\Pi})
  \mat{U}^{\rm T} \mat{N} = \mat{U} \mat{\Pi} \mat{U}^{\rm T} \mat{N}.
\end{equation}
Since $\mat{U}^{\rm T} \mat{U}_{f} \ne \mat{0}$ and $\mat{U}^{\rm T}
\mat{U}_{n} \ne \mat{0}$ in general, 
$\mat{F}^{\text{mix}}_{\rm PCA/SVD}$ and $\mat{N}^{\text{loss}}_{\rm PCA/SVD}$ are non-zero, i.e.\ in the standard PCA/SVD method, there are signal loss and foreground mixing, if the conditions in \S\ref{S:cond} are not met. 

We define the {\it recovery error} as the difference between the true signal and the recovered signal, $\Delta \mat{N} = \mat{N} - \hat{\mat{N}}$. For PCA/SVD, the recovery error is 
\begin{equation}
    \Delta \mat{N}_{\rm PCA/SVD} = \mat{N}^{\text{loss}}_{\rm PCA/SVD} - \mat{F}^{\text{mix}}_{\rm PCA/SVD}\,.
\end{equation}



\section{Singular Vector Projection} \label{S:sigvec}

In what we call the ``Singular Vector Projection'' method, we propose to exploit the information of the singular vectors of the foregrounds, $\mat{U}_f$ and/or $\mat{V}_f$, if it is known {\it a priori}, to improve the accuracy of foreground subtraction. We propose the new estimators in \S\ref{S:u} and \S\ref{S:rf}, provide a pedagogical example in \S\ref{S:se}, and discuss the feasibility of obtaining {\it a priori} information of $\mat{U}_f$ and/or $\mat{V}_f$ in \S\ref{S:mv}. 

\subsection{The SVP Estimators}\label{S:u}


We propose four estimators for the data $\mat{D}$, as follows. 
\begin{align}
  \mat{N}_{\text{L}} &= \mat{D} - \mat{U}_{f} \mat{U}_{f}^{\rm T} \mat{D} \,,  \label{eq:SVP-D-L} \\
  \mat{N}_{\text{R}} &= \mat{D} - \mat{D} \mat{V}_{f} \mat{V}_{f}^{\rm T} \,, \\
  \mat{N}_{\text{B}} &= \mat{D} - \mat{U}_{f} \mat{U}_{f}^{\rm T} \mat{D} \mat{V}_{f} \mat{V}_{f}^{\rm T} \,,\\
  \mat{N}_{\text{D}} &= \mat{D} - \mat{U}_{f} (\mat{U}_{f}^{\rm T} \mat{D} \mat{V}_{f})_{\text{diag}} \mat{V}_{f}^{\rm T} \,.\label{eq:SVP-D-D}
\end{align}
Here, $\mathcal{O}_{\text{diag}}$ denotes a diagonal matrix in which its elements are the same as the diagonal elements of the matrix $\mathcal{O}$. 

If only the left (right) singular vector of the foregrounds, $\mat{U}_f$ ($\mat{V}_f$), is known {\it a priori}, then the estimator $\mat{N}_{\text{L}}$ ($\mat{N}_{\text{R}}$) can be applied. If both left and right singular vectors of the foregrounds are known {\it a priori}, then the estimators $\mat{N}_{\text{B}}$ and $\mat{N}_{\text{D}}$ can be applied. The subscripts ``L'', ``R'', ``B'', ``D'' stand for ``left'', ``right'', ``both'', and ``diagonal'', respectively. 

It is straightforward to prove the following results, 
\begin{align}
  \mat{N}_{\text{L}} &=  \mat{N} - \mat{U}_{f} \mat{U}_{f}^{\rm T} \mat{N}, \label{eq:SVP-L} \\
  \mat{N}_{\text{R}} &=  \mat{N} - \mat{N} \mat{V}_{f} \mat{V}_{f}^{\rm T}, \label{eq:SVP-R} \\
  \mat{N}_{\text{B}} &=  \mat{N} - \mat{U}_{f} \mat{U}_{f}^{\rm T} \mat{N} \mat{V}_{f} \mat{V}_{f}^{\rm T}, \label{eq:SVP-B} \\
  \mat{N}_{\text{D}} &=  \mat{N} - \mat{U}_{f} (\mat{U}_{f}^{\rm T} \mat{N} \mat{V}_{f})_{\text{diag}} \mat{V}_{f}^{\rm T}. \label{eq:SVP-D}
\end{align}
This shows that these estimators can project out the foregrounds $\mat{F}$ completely, i.e.\ no foreground mixing. In fact, these are the only four estimators that meet this requirement. This is an advantage against the blind PCA/SVD method, because given that the foreground is several orders of magnitude stronger than the signal, even a small residual foreground mixing can result in large recovery error. 
The proof of Eqs.~(\ref{eq:SVP-L})---(\ref{eq:SVP-B}) uses the identity $\mat{U}_{f} \mat{U}_{f}^{\rm T} \mat{F} = \mat{F} \mat{V}_{f} \mat{V}_{f}^{\rm T} = \mat{F}$, which is due to $\mat{U}_f^{\rm T} \mat{U}_f = \mat{I}$ and $\mat{V}_f^{\rm T} \mat{V}_f = \mat{I}$. The proof of Eq.~(\ref{eq:SVP-D}) uses the identity $(\mat{U}_{f}^{\rm T} \mat{F} \mat{V}_{f})_{\text{diag}} = (\mat{S}_f)_{\text{diag}} = \mat{S}_f$, because $\mat{S}_f$ is diagonal.

The recovery error and signal loss for these estimators are as follows, respectively. 
\begin{align}
  \Delta \mat{N}_{\text{L}} = \mat{N}_{\text{L}}^{\text{loss}} &= \mat{U}_{f} \mat{U}_{f}^{\rm T} \mat{N}\,, \notag \\
  \Delta \mat{N}_{\text{R}} = \mat{N}_{\text{R}}^{\text{loss}} &= \mat{N} \mat{V}_{f} \mat{V}_{f}^{\rm T}\,, \notag \\
  \Delta \mat{N}_{\text{B}} = \mat{N}_{\text{B}}^{\text{loss}} &= \mat{U}_{f} \mat{U}_{f}^{\rm T} \mat{N}
                              \mat{V}_{f} \mat{V}_{f}^{\rm T}\,, \notag \\
  \Delta \mat{N}_{\text{D}} = \mat{N}_{\text{D}}^{\text{loss}} &= \mat{U}_{f} (\mat{U}_{f}^{\rm T} \mat{N}
                           \mat{V}_{f})_{\text{diag}} \mat{V}_{f}^{\rm T}\,. \label{eq:4loss}
\end{align}

If the condition for the left singular vectors $\mat{U}_{f}^{\rm T} \mat{U}_{n} =
\mat{0}$ is met, then $\Delta \mat{N}_{\text{L}} = \Delta \mat{N}_{\text{B}} = \Delta \mat{N}_{\text{D}}  = \mat{0}$, and if the condition for the right singular vectors $\mat{V}_{n}^{\rm T} \mat{V}_{f} = \mat{0}$ is met, then $\Delta \mat{N}_{\text{R}} = \Delta \mat{N}_{\text{B}} = \Delta \mat{N}_{\text{D}}  = \mat{0}$. Clearly, using the estimators $\mat{N}_{\text{B}}$ and $\mat{N}_{\text{D}}$ with the information of both left and right singular vectors is more likely to get smaller recovery error than using either $\mat{N}_{\text{L}}$ or $\mat{N}_{\text{R}}$ with only the left or right singular vector. In fact, roughly speaking, the ``magnitude'' of recovery error for these estimators have the following relation: $\Delta \mat{N}_{\text{D}} \le \Delta \mat{N}_{\text{B}} \le \Delta \mat{N}_{\text{L}}$ and $\Delta \mat{N}_{\text{D}} \le \Delta \mat{N}_{\text{B}} \le \Delta \mat{N}_{\text{R}}$. We leave it to Appendix~\ref{S:csl} to give an accurate definition of their ``magnitudes'' and the proof of these relations. Note that these inequality relations are valid for the whole matrix, and do not necessarily hold for each individual frequency bin. 
\subsection{SVP with Incomplete Singular Vectors} \label{S:rf}

In \S\ref{S:u}, we implicitly assume that the left and/or right singular vectors of the foregrounds, $\mat{U}_f \in \mathbb{R}^{n \times k}$ and $\mat{V}_f \in \mathbb{R}^{p \times k}$, for {\it all} $k$ modes can be well modeled or measured {\it a priori}. In practice, this might not be satisfied. 

Consider a relaxed condition in which the left and/or right singular vectors of the foregrounds for only a small number of modes which correspond to the $l$ largest singular values can be modeled {\it a priori}, labeled as $\mat{U}_{f_{1}} \in \mathbb{R}^{n \times l}$ and $\mat{V}_{f_{1}} \in \mathbb{R}^{p \times l}$, and the other $k-l$ singular vectors of the foregrounds, labeled as $\mat{U}_{f_2}
\in \mathbb{R}^{n \times (k-l)}$ and $\mat{V}_{f_2} \in\mathbb{R}^{p \times (k-l)}$, are not known. The total foregrounds can be written as the sum of two parts, $\mat{F} = \mat{F}_{1} + \mat{F}_{2}$, where 
\begin{align}
  \mat{F}_{1} &= \mat{U}_{f_1} \mat{S}_{f_1} \mat{V}^{\rm T}_{f_1}, \notag \\
  \mat{F}_{2} &= \mat{U}_{f_2} \mat{S}_{f_2} \mat{V}^{\rm T}_{f_2}. \label{eq:F12}
\end{align}
Here, $\mat{S}_{f_1}\in\mathbb{R}^{l \times l}$ is the diagonal matrix of the $l$ largest singular values, and $\mat{S}_{f_2} \in\mathbb{R}^{(k-l) \times (k-l)}$ is the diagonal matrix of the rest $k-l$ singular values. We assume that the former is significantly larger than the latter. 

In this case with incomplete information of singular vectors of the foregrounds, the estimators for the data $\mat{D}$ are as follows. 
\begin{align}
  \mat{N}_{\text{L}} &= \mat{D} - \mat{U}_{f_1} \mat{U}_{f_1}^{\rm T} \mat{D} \,, \\
  \mat{N}_{\text{R}} &= \mat{D} - \mat{D} \mat{V}_{f_1} \mat{V}_{f_1}^{\rm T} \,, \\
  \mat{N}_{\text{B}} &= \mat{D} - \mat{U}_{f_1} \mat{U}_{f_1}^{\rm T} \mat{D} \mat{V}_{f_1} \mat{V}_{f_1}^{\rm T} \,,\\
  \mat{N}_{\text{D}} &= \mat{D} - \mat{U}_{f_1} (\mat{U}_{f_1}^{\rm T} \mat{D} \mat{V}_{f_1})_{\text{diag}} \mat{V}_{f_1}^{\rm T} \,.
\end{align}

Similar to the results in \S\ref{S:u}, it is straightforward to prove the following results, 
\begin{align}
  \mat{N}_{\text{L}} &=  \mat{N} - \mat{U}_{f_1} \mat{U}_{f_1}^{\rm T} \mat{N} + \mat{F}_{2} \,, \label{eqn:L1}  \\
  \mat{N}_{\text{R}} &=  \mat{N} - \mat{N} \mat{V}_{f_1} \mat{V}_{f_1}^{\rm T} + \mat{F}_{2} \,,  \\
  \mat{N}_{\text{B}} &=  \mat{N} - \mat{U}_{f_1} \mat{U}_{f_1}^{\rm T} \mat{N} \mat{V}_{f_1} \mat{V}_{f_1}^{\rm T}+\mat{F}_{2} \,,  \\
  \mat{N}_{\text{D}} &=  \mat{N} - \mat{U}_{f_1} (\mat{U}_{f_1}^{\rm T} \mat{N} \mat{V}_{f_1})_{\text{diag}} \mat{V}_{f_1}^{\rm T} + \mat{F}_{2} \,.\label{eqn:D1}
\end{align}
This shows that these estimators project out the major foreground component $\mat{F}_{1}$, but the unknown, minor foreground component $\mat{F}_{2}$ is left as the residual foreground mixing, $\mat{F}^{\text{mix}} = \mat{F}_{2}$. 
The proof of Eqs.~(\ref{eqn:L1})---(\ref{eqn:D1}) uses the orthogonality relation $\mat{U}^{\rm T}_{f_1} \mat{U}_{f_2} = \mat{0}$ and $\mat{V}^{\rm T}_{f_1} \mat{V}_{f_2} = \mat{0}$, and the identities $\mat{U}_{f_1} \mat{U}^{\rm T}_{f_1} \mat{F}_1 =
\mat{F}_1 \mat{V}_{f_1} \mat{V}_{f_1}^{\rm T} = \mat{U}_{f_1} \mat{U}^{\rm T}_{f_1} \mat{F}_1 \mat{V}_{f_1} \mat{V}_{f_1}^{\rm T} = \mat{U}_{f_1} (\mat{U}^{\rm T}_{f_1} \mat{F}_1 \mat{V}_{f_1})_{\text{diag}} \mat{V}_{f_1}^{\rm T} = \mat{F}_{1}$. 

The recovery errors for these estimators are as follows. 
\begin{align}
  \Delta \mat{N}_{\text{L}} &= \mat{U}_{f_1} \mat{U}_{f_1}^{\rm T} \mat{N} - \mat{F}_{2}\,, \notag \\
  \Delta \mat{N}_{\text{R}} &= \mat{N} \mat{V}_{f_1} \mat{V}_{f_1}^{\rm T} - \mat{F}_{2}\,, \notag \\
  \Delta \mat{N}_{\text{B}} &= \mat{U}_{f_1} \mat{U}_{f_1}^{\rm T} \mat{N} \mat{V}_{f_1} \mat{V}_{f_1}^{\rm T}-\mat{F}_{2} \,, \notag \\
  \Delta \mat{N}_{\text{D}} &= \mat{U}_{f_1} (\mat{U}_{f_1}^{\rm T} \mat{N} \mat{V}_{f_1})_{\text{diag}} \mat{V}_{f_1}^{\rm T} - \mat{F}_{2} \,. \label{eq:svpinc}
\end{align}

Finally, we note that there is actually a freedom of normalization in the singular vectors. However, our SVP estimators contain the singular vectors that are always in pairs, such as $\mat{U}_{f} \mat{U}_{f}^{\rm T}$ and $\mat{V}_{f} \mat{V}_{f}^{\rm T}$, so the estimators are independent of this normalization in the singular vectors. Also, the estimators do not depend on the eigenvalues $\mat{\Lambda}_{f}$ of the covariance matrix, so the estimators are not affected by the overall magnitude of the covariance matrix. In other words, even if the covariance matrix of the foregrounds was biased in overall amplitude from observations, the SVP estimators would not be affected.

\begin{figure*}
  \centering
  \includegraphics[width=0.4\textwidth]{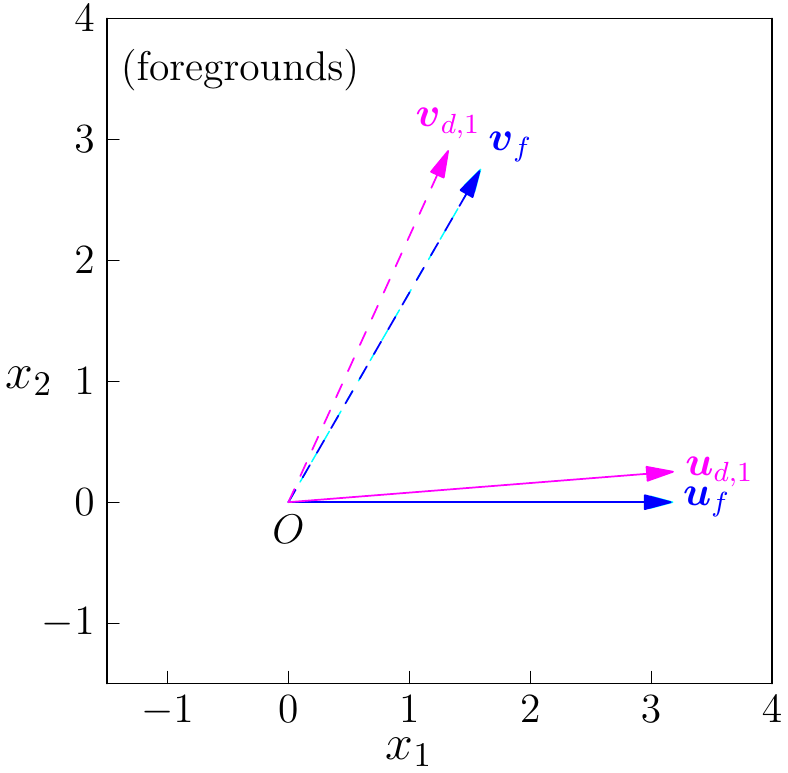}
  \includegraphics[width=0.4\textwidth]{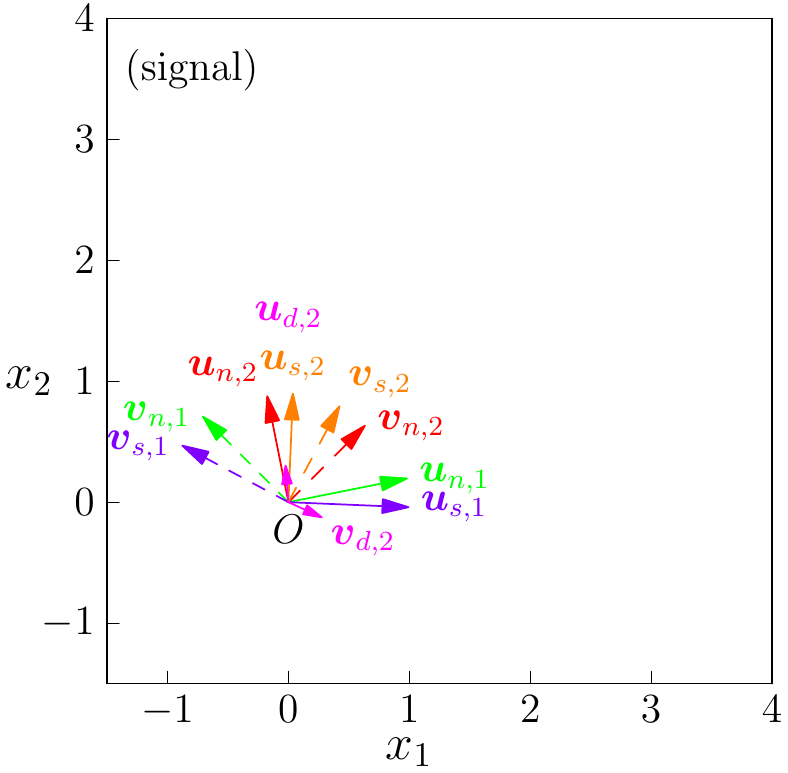}
  \caption{Illustration of foreground subtraction with SVP: foregrounds (left panel) and signal (right panel). Specifically, we show the left (right) singular vectors $\sqrt{s_{f}}\vec{u}_{f}$ ($\sqrt{s_{f}}\vec{v}_{f}$) of the true foregrounds $\mat{F} = s_{f} \vec{u}_{f} \vec{v}_f^{\rm T}$ (blue), and the left (right) singular vectors $\sqrt{s_{n,1}}\vec{u}_{n,1}$ and $\sqrt{s_{n,2}}\vec{u}_{n,2}$ ($\sqrt{s_{n,1}}\vec{v}_{n,1}$ and $\sqrt{s_{n,2}}\vec{v}_{n,2}$) of the signal $\mat{N} = s_{n,1} \vec{u}_{n,1} \vec{v}_{n,1}^{\rm T} + s_{n,2} \vec{u}_{n,2} \vec{v}_{n,2}^{\rm T}$ (green and red). We use the PCA/SVD method to decompose the image $\mat{D} = s_{d,1} \vec{u}_{d,1} \vec{v}_{d,1}^{\rm T} + s_{d,2}\vec{u}_{d,2} \vec{v}_{d,2}^{\rm T}$, in which the first (second) mode is identified as the estimated foregrounds (recovered signal). We plot their left (right) singular vectors $\sqrt{s_{d,1}} \vec{u}_{d,1}$ and $\sqrt{s_{d,2}} \vec{u}_{d,2}$ ($\sqrt{s_{d,1}} \vec{v}_{d,1}$ and $\sqrt{s_{d,2}} \vec{v}_{d,2}$), respectively (magenta). We also apply the SVP estimator (``Both'' and ``Diagonal'' which are the same in this simple example) to estimate the foregrounds $\mat{F}_{\rm B/D}  = s_{df} \vec{u}_{f} \vec{v}_f^{\rm T}$ and recover the signal $\mat{N}_{\rm B/D} = s_{s,1} \vec{u}_{s,1}
\vec{v}_{s,1}^{\rm T} + s_{s,2} \vec{u}_{s,2} \vec{v}_{s,2}^{\rm T}$, and plot the left (right) singular vectors $\sqrt{s_{df}} \vec{u}_{f}$ ($\sqrt{s_{df}} \vec{v}_{f}$) of the foregrounds (cyan), and $\sqrt{s_{s,1}} \vec{u}_{s,1}$ and $\sqrt{s_{s,2}} \vec{u}_{s,2}$ ($\sqrt{s_{s,1}} \vec{v}_{s,1}$ and $\sqrt{s_{s,2}} \vec{v}_{s,2}$) of the recovered signal (purple and orange). Note that the cyan line overlaps with the blue line in this simple example, albeit slightly longer. For brevity, the rescaling factors are omitted in the legends of these vectors. 
}
  \label{fig:uv}
\end{figure*}

\begin{figure*}
  \centering
  \includegraphics[width=0.45\textwidth]{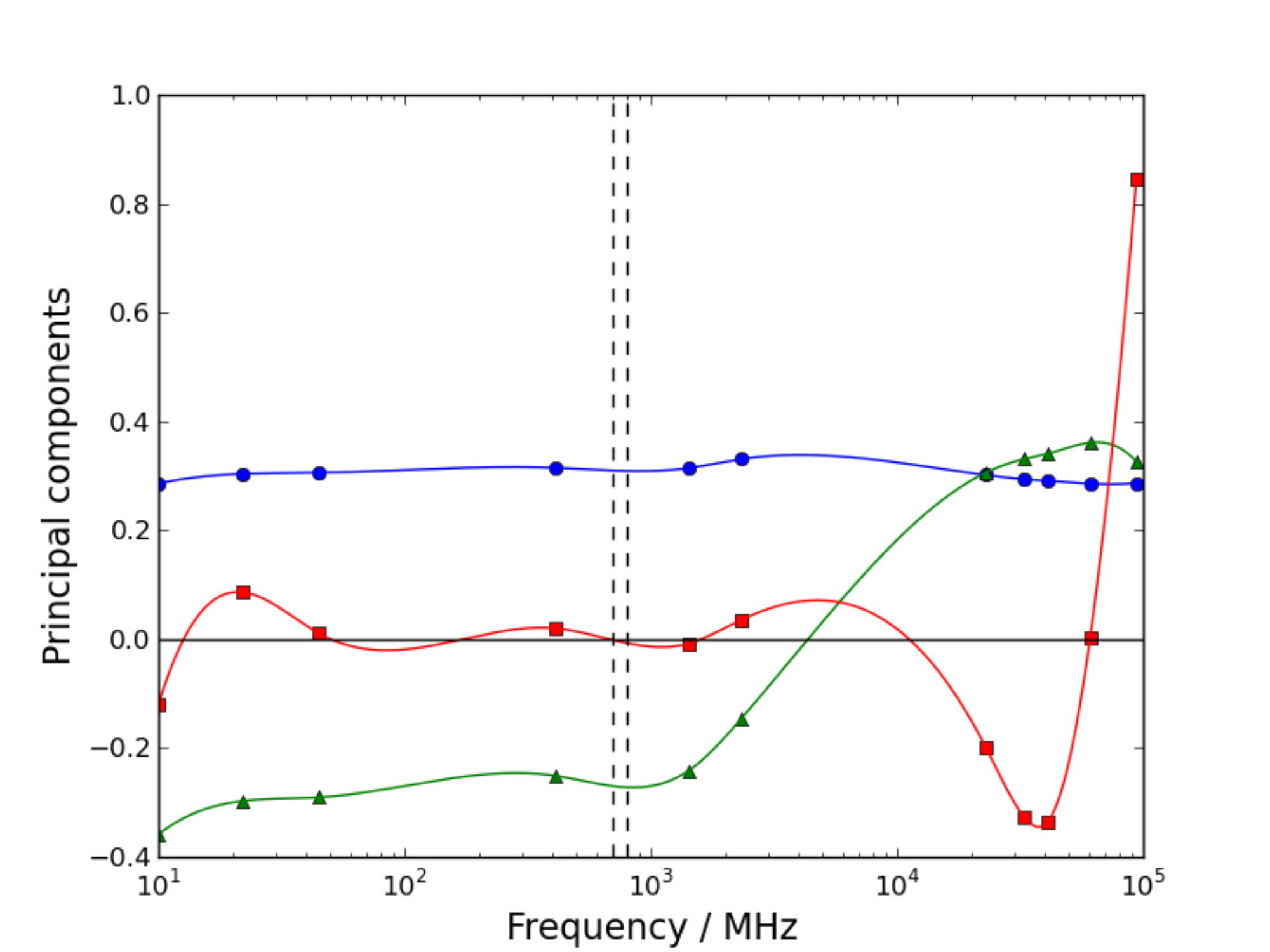}
  \includegraphics[width=0.45\textwidth]{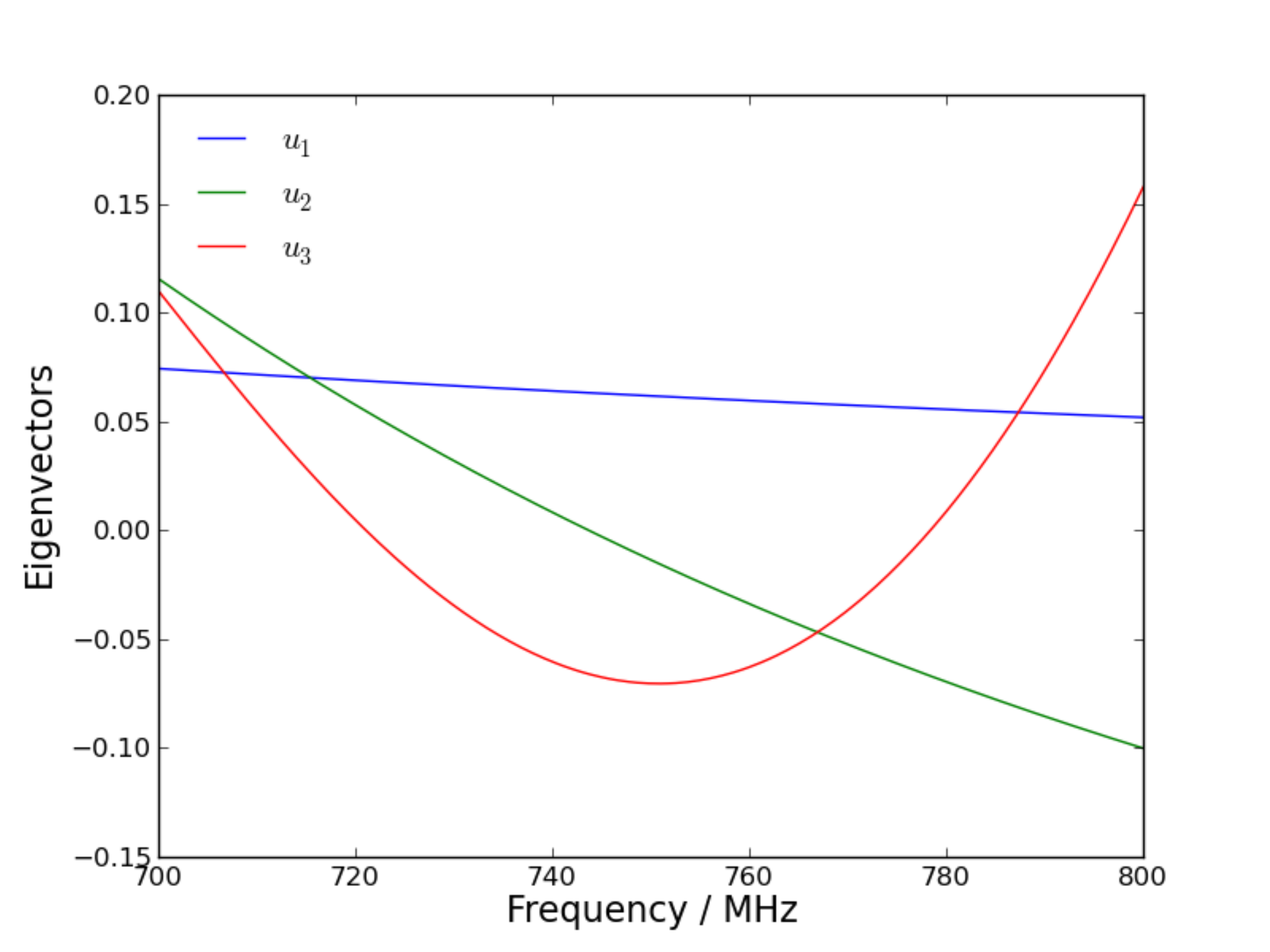}
  \caption{(Left) Three principal components that correspond to the largest three eigenvalues of the correlation matrix (from the large to small eigenvalue, in blue, green and red, respectively), using the sky region with data at all 11 frequencies (shown in dots). We fit the frequency dependence with a cubic spline in the frequency band of $700 - 800$~MHz as enveloped by two black dashed lines. (Right) Three left singular vectors of the foregrounds corresponding to the largest three eigenvalues of the frequency covariance matrix that is constructed from the sky map of predicted foreground emission in the frequency band of $700 - 800$~MHz. The eigenvectors are unitless and normalized to unity. 
  }
  \label{fig:fguvec}
\end{figure*}

\subsection{Pedagogical Example} \label{S:se}

In this subsection, we use a pedagogical example to illustrate the SVP. Consider a simple case where the image $\mat{D}$ has two frequency bins and two pixels. Suppose the foregrounds is a rank-one matrix, $\mat{F}
= s_{f} \vec{u}_{f} \vec{v}_f^{\rm T}$ for simplicity, and the signal is a rank-two matrix, $\mat{N} =
s_{n,1} \vec{u}_{n,1} \vec{v}_{n,1}^{\rm T} + s_{n,2} \vec{u}_{n,2} \vec{v}_{n,2}^{\rm T}$. Here $\vec{u}_{f}$ ($\vec{v}_{f}$) is the left (right) singular vector of $\mat{F}$; 
$\vec{u}_{n,1}$ and $\vec{u}_{n,2}$ ($\vec{v}_{n,1}$ and $\vec{v}_{n,2}$) are the left (right) singular vectors of $\mat{N}$, and perpendicular to each other, i.e.\ $\vec{u}_{n,1} \perp \vec{u}_{n,2}$ and $\vec{v}_{n,1} \perp \vec{v}_{n,2}$. They are all unit vectors of the size $2\times 1$. In this example, we set $s_{f} = 10.0$,
$s_{n,1} = 1.0$ and $s_{n,2} = 0.8$, so that the magnitude of $\mat{F}$ is about ten times larger than $\mat{N}$. For such a simple case, vectors of the size $2\times 1$ can be visualized in a 2D plane, so we plot $\mat{F}$ by the rescaled vectors $\sqrt{s_{f}} \vec{u}_{f}$ and $\sqrt{s_{f}} \vec{v}_{f}$, and plot $\mat{N}$ by the rescaled vectors 
$\sqrt{s_{n,1}} \vec{u}_{n,1}$, $\sqrt{s_{n,2}} \vec{u}_{n,2}$, 
$\sqrt{s_{n,1}} \vec{v}_{n,1}$, and $\sqrt{s_{n,2}} \vec{v}_{n,2}$ in Figure~\ref{fig:uv}.
Similar rescaling will be applied when we refer to the ``magnitude'' of a vector in this subsection. 

In the standard SVD analysis, the image is decomposed as  
$\mat{D} = s_{d,1} \vec{u}_{d,1} \vec{v}_{d,1}^{\rm T} + s_{d,2}\vec{u}_{d,2} \vec{v}_{d,2}^{\rm T}$. Here, $s_{d,1} > s_{d,2}$, and $\vec{u}_{d,1}\perp \vec{u}_{d,2}$, $\vec{v}_{d,1} \perp \vec{v}_{d,2}$. In this decomposition, the first term $s_{d,1} \vec{u}_{d,1} \vec{v}_{d,1}^{\rm T}$ is identified and thus removed as the foregrounds, but Figure~\ref{fig:uv} shows that the direction of the left (right) singular vector $\vec{u}_{d,1}$ ($\vec{v}_{d,1}$) is different from that of the left (right) singular vectors $\vec{u}_{f}$ ($\vec{v}_{f}$) of the true foregrounds $\mat{F}$. On the other hand, the second term $s_{d,2}\vec{u}_{d,2} \vec{v}_{d,2}^{\rm T}$ of the SVD decomposition is identified as the recovered signal $\mat{N}_{\rm PCA/SVD}$, but it is rank-one in this case, while the true signal $\mat{N}$ is rank-two. Also, Figure~\ref{fig:uv} shows that the magnitude of $\mat{N}_{\rm PCA/SVD}$ is much smaller than $\mat{N}$, which means a significant signal loss in the SVD foreground removal. 

Now, assuming that we know $\vec{u}_{f}$ and $\vec{v}_{f}$, as discussed in Sec.~\ref{S:u}, we can project out the foregrounds in four ways. The optimal estimators are $\mat{N}_{\rm B}$ and/or $\mat{N}_{\rm D}$ that exploit the information of both $\vec{u}_{f}$ and $\vec{v}_{f}$. For the pedagogical example considered herein, the two estimators give the same results because $\vec{u}_{f}^{\rm T} \mat{D} \vec{v}_{f}$ is a $1 \times 1$ matrix, i.e. a number, in this example, so $\vec{u}_{f}^{\rm T} \mat{D} \vec{v}_{f} = (\vec{u}_{f}^{\rm T} \mat{D} \vec{v}_{f})_{\text{diag}}$. We can rewrite the estimated foregrounds as $\mat{F}_{\rm B/D}  = s_{df} \vec{u}_{f} \vec{v}_f^{\rm T}$, where we define $s_{df} \equiv \vec{u}_f^{\rm T} \mat{D} \vec{v}_{f}$ for this case. The left (right) singular vector $\sqrt{s_{df}} \vec{u}_{f}$ ($\sqrt{s_{df}} \vec{v}_{f}$) of the estimated foregrounds $\mat{F}_{\rm B/D}$ has the same direction as the left (right) singular vector $\sqrt{s_{f}}
\vec{u}_{f}$ ($\sqrt{s_{f}} \vec{v}_{f}$) of the true foregrounds $\mat{F}$. In principle, the magnitude of the former is slightly larger than the latter, because $s_{df} > s_{f}$, but in this simple example, their magnitudes are very close, as shown in Figure~\ref{fig:uv}. This implies that
the subtracted foregrounds contain the full foregrounds $\mat{F}$ and some amount of signal. So the recovered signal, $\mat{N}_{\rm B/D} = \mat{D} - \mat{F}_{\rm B/D}$ is free from contamination of any residual foregrounds, but suffers from some amount of signal loss. Also, $\mat{N}_{\rm B/D} = s_{s,1} \vec{u}_{s,1}
\vec{v}_{s,1}^{\rm T} + s_{s,2} \vec{u}_{s,2} \vec{v}_{s,2}^{\rm T}$ is still a rank-two matrix. Figure~\ref{fig:uv} shows that the recovered signal $\mat{N}_{\rm B/D}$ is close to the true signal $\mat{N}$. This simple example clearly shows that using the information of the left and right singular vectors of the foregrounds can help improve the performance of foreground subtraction.

\subsection{Modeling the Singular Vectors} \label{S:mv}

The left and right singular vectors of the foregrounds, $\mat{U}_{f}$ and $\mat{V}_f$, can be obtained by solving for the eigen-decomposition of the frequency and pixel covariance matrix, respectively, i.e.\ $\mat{F} \mat{F}^{\rm T} = \mat{U}_{f} \mat{\Lambda}_{f} \mat{U}_{f}^{\rm T}$, and $\mat{F}^{\rm T} \mat{F} = \mat{V}_{f} \mat{\Lambda}_{f} \mat{V}_{f}^{\rm T}$, where $\mat{\Lambda}_{f} = \mat{S}_{f}^2$. Our assumption is that the covariance matrix of the foregrounds can be estimated from modeling or observations {\it a priori}. In practice, we may keep only the singular vectors with positive eigenvalues, and drop all modes with nearly zero eigenvalues. 

We follow \citet{Costa2008}, which is essentially a PCA method, to estimate the foregrounds. We summarize the approach below, and refer interested readers to \citet{Costa2008} for details. 
Typically, we have the dataset from a number of available surveys that altogether can cover the full sky, but their overlapping sky coverage is only a patch of sky with $n_{\text{pix}}$ pixels with data at all $n_f$ frequencies. For the data subset of the overlapping sky map $\mat{y} \in \mathbb{R}^{n_f \times  n_{\rm pix}}$, we begin by estimating its matrix of second moments of the size $n_{f} \times n_{f}$ which is essentially the normalized covariance matrix $ \mat{C} \equiv \mat{y} \mat{y}^{\rm T}/n_{\text{pix}} $, and then estimating the correlation matrix $\mathcal{R}_{ij} \equiv C_{ij}/\sigma_{i} \sigma_{j}$ that corresponds to the dimensionless correlation coefficients between all pairs of frequencies. Here $\sigma_{i} \equiv C_{ii}^{1/2}$ is the rms fluctuations at each frequency. 
We then perform a standard eigenvalue decomposition to diagonalize
the matrix $\mat{\mathcal{R}}$ as $\mat{\mathcal{R}} = \mat{P} \mat{\Lambda} \mat{P}^{\rm T}$, 
where $\mat{P}$ is an orthogonal matrix in which the columns are the
eigenvectors (principal components) and $\mat{\Lambda}$ is a
diagonal matrix containing the corresponding eigenvalues. 
We determine the first $n_{c}$ principal components that best
approximate the data using the overlapping sky region with data at all
$n_{f}$ frequencies according to some accuracy criteria (see \citealt{Costa2008}). 
We then fit for the $n_c$ principal component maps (the matrix product of the $n_c$ principal components and the normalized maps, i.e. the input maps rescaled to have rms fluctuations of unity at each frequency) across the entire sky pixel by pixel by using the normalized input maps that have data for that pixel. To predict sky maps at other frequencies, we can further fit the frequency dependence of both $\log{\sigma_{i}}$ and each of the best $n_{c}$ principal components with a cubic spline as a function of $\log{\nu}$, and use them and the fitted $n_c$ principal component maps to reconstruct the sky maps at the required frequencies. 
Here we implicitly assume that these are smooth, slowly varying functions (as shown in Figure~\ref{fig:fguvec}). With this machinery, we can construct a map of the foreground emission in the target frequency range with $n$ frequency bins in the target patches of sky with $p$ pixels, from which we can estimate the frequency (pixel) covariance matrix, and subsequently the left (right) singular vectors of the foregrounds by an eigenvalue decomposition of the covariance matrix. Note that this singular vector modeling process can be improved as more, higher-quality survey data at different frequencies become available. Advanced techniques presented in \citet{Zheng2017} may be also applied to better
account for different survey data that has non-overlapping regions. 

As an example, we apply this method for modeling the singular vectors to the Tianlai array \citep{Chen2012,Xu2015}, which operates in the frequency band between 700 and 800 MHz with 256 frequency bins. We use the survey data at the same 11 frequencies (taken from the {\tt PyGSM}\footnote{\url{https://github.com/telegraphic/PyGSM}} package) as used in \citet{Costa2008}. 
The first three principal components with the largest three eigenvalues are plotted in the left panel of Figure~\ref{fig:fguvec}. From this, we can predict the frequency dependence at the frequencies between 700 and 800 MHz where no survey data is available, by fitting each principal component with a cubic spline as a function of $\log{\nu}$. Figure~\ref{fig:fguvec} demonstrates that this fitting works well, because the
frequency dependencies of these functions are indeed smooth and slowly varying in this frequency band. 
From the sky map of predicted foreground emission in this frequency band, we estimate the frequency (pixel) covariance matrix and solve for the left (right) singular vector of the foregrounds. As an illustration, we show the three left singular vectors of the foregrounds corresponding to the largest three eigenvalues in the right panel of Figure~\ref{fig:fguvec}. 

\subsection{What if sky model is biased?} \label{S:mu}

In practical analysis, it is likely that we employ a biased foreground model. In this subsection, we discuss how uncertainties in the sky model impact the foreground subtraction with SVP. 

Suppose we employ a foreground model $\mat{F'}$, which is biased from the true foreground $\mat{F}$ by $\mat{F} = \mat{F'} + \Delta \mat{F}$. We assume the foregrounds are decomposed using SVD by $\mat{F} = \mat{U}_f \mat{S}_f \mat{V}_f^{\rm T}$,  $\mat{F}' = \mat{U}_{f'} \mat{S}_{f'} \mat{V}_{f'}^{\rm T}$ and $\Delta \mat{F} = \mat{U}_{\Delta f} \mat{S}_{\Delta f} \mat{V}_{\Delta f}^{\rm T}$. 
With this sky model, we apply the biased singular vectors $\mat{U}_{f'}$ and $\mat{V}_{f'}$, instead of $\mat{U}_f$ and $\mat{V}_f$, to the observational data $\mat{D}$ using Eqs.~(\ref{eq:SVP-D-L})---(\ref{eq:SVP-D-D}). Since $\mat{D}= \mat{F} + \mat{N}= \mat{F}' + (\mat{N} + \Delta \mat{F})$, we can simply replace $\mat{N}$ in Eqs.~(\ref{eq:SVP-L})---(\ref{eq:SVP-D}) by $\mat{N} + \Delta \mat{F}$, and find that
\begin{align}
  \mat{N}'_{\text{L}} &=  (\mat{N} - \mat{U}_{f'} \mat{U}_{f'}^{\rm T} \mat{N}) + (\Delta  \mat{F} - \mat{U}_{f'} \mat{U}_{f'}^{\rm T} \Delta \mat{F}), \label{eq:SVP-Lp} \\
  \mat{N}'_{\text{R}} &=  (\mat{N} - \mat{N} \mat{V}_{f'} \mat{V}_{f'}^{\rm T}) + (\Delta \mat{F} - \Delta \mat{F} \mat{V}_{f'} \mat{V}_{f'}^{\rm T}), \label{eq:SVP-Rp} \\
  \mat{N}'_{\text{B}} &=  (\mat{N} - \mat{U}_{f'} \mat{U}_{f'}^{\rm T} \mat{N} \mat{V}_{f'} \mat{V}_{f'}^{\rm T}) \notag \\
  &+ (\Delta \mat{F} - \mat{U}_{f'} \mat{U}_{f'}^{\rm T} \Delta \mat{F} \mat{V}_{f'} \mat{V}_{f'}^{\rm T}), \label{eq:SVP-Bp} \\
  \mat{N}'_{\text{D}} &=  (\mat{N} - \mat{U}_{f'} (\mat{U}_{f'}^{\rm T} \mat{N} \mat{V}_{f'})_{\text{diag}} \mat{V}_{f'}^{\rm T}) \notag \\
  &+ (\Delta \mat{F} - \mat{U}_{f'} (\mat{U}_{f'}^{\rm T} \Delta \mat{F} \mat{V}_{f'})_{\text{diag}} \mat{V}_{f'}^{\rm T}). \label{eq:SVP-Dp}
\end{align}
The first bracketed terms in Eqs.~(\ref{eq:SVP-Lp})---(\ref{eq:SVP-Dp}) should approximate to the corresponding terms given in Eqs.~(\ref{eq:SVP-L})---(\ref{eq:SVP-D}), if the foreground model $\mat{F}'$ is not too much biased from the true foreground $\mat{F}$. Regarding the second  bracketed terms (i.e.~foreground mixing terms) in Eqs.~(\ref{eq:SVP-Lp})---(\ref{eq:SVP-Dp}), we first consider two extreme cases.

(i) The singular vectors of the offset $\Delta \mat{F}$ are aligned with those of the foreground model $\mat{F'}$, i.e.~$\mat{U}_{f'} \mat{U}_{f'}^{\rm T} \mat{U}_{\Delta f} = \mat{U}_{\Delta f}$, $\mat{V}^{\rm T}_{\Delta f} \mat{V}_{f'} \mat{V}_{f'}^{\rm T} = \mat{V}^{\rm T}_{\Delta f}$, which means that $\mat{U}_{f'} \mat{U}_{f'}^{\rm T} \Delta \mat{F} = \Delta \mat{F}$, and $\Delta \mat{F} \mat{V}_{f'} \mat{V}_{f'}^{\rm T} = \Delta \mat{F}$, and therefore $\mat{U}_{f'} \mat{U}_{f'}^{\rm T} \Delta \mat{F} \mat{V}_{f'} \mat{V}_{f'}^{\rm T} = \Delta \mat{F}$. This means that the second bracketed terms in Eqs.~(\ref{eq:SVP-Lp})---(\ref{eq:SVP-Bp}) vanish (but the second bracketed term in Eq.~(\ref{eq:SVP-Dp}) is not necessarily zero).
 
A stronger example is the case where the foreground model is only off the true foreground by an overall factor, i.e.~$\mat{F}' = (1 - \alpha) \mat{F} = \mat{U}_f [(1 - \alpha) \mat{S}_f] \mat{V}_f^{\rm T}$ and $\Delta \mat{F} = \alpha \mat{F} = \mat{U}_f (\alpha \mat{S}_f) \mat{V}_f^{\rm T}$, where $\alpha$ is a real number. This means that the singular vectors of $\mat{F'}$, $\Delta \mat{F}$ and $\mat{F}$ are just the same, and the overall bias factor of foregrounds only affects the singular values. In this case, $\mat{U}_{f'} \mat{U}_{f'}^{\rm T} \Delta \mat{F} = \Delta \mat{F} \mat{V}_{f'} \mat{V}_{f'}^{\rm T} = \mat{U}_{f'} \mat{U}_{f'}^{\rm T} \Delta \mat{F} \mat{V}_{f'} \mat{V}_{f'}^{\rm T} = \mat{U}_{f'} (\mat{U}_{f'}^{\rm T} \Delta \mat{F} \mat{V}_{f'})_{\text{diag}} \mat{V}_{f'}^{\rm T} = \Delta \mat{F}$, and as such, all second bracketed terms in Eqs.~(\ref{eq:SVP-Lp})---(\ref{eq:SVP-Dp}) vanish, and we get exactly the same results as given in Eqs.~(\ref{eq:SVP-L})---(\ref{eq:SVP-D}). This just reflects the fact that the overall magnitude of the foreground does not impact the performance of the SVP method.

(ii) The singular vectors of the offset $\Delta \mat{F}$ are orthogonal to those of the foreground model $\mat{F'}$, i.e.~$\mat{U}_{f'}^{\rm T} \mat{U}_{\Delta f} = \mat{0}$, $\mat{V}^{\rm T}_{\Delta f} \mat{V}_{f'}  = \mat{0}$, which means that $\mat{U}_{f'} \mat{U}_{f'}^{\rm T} \Delta \mat{F} = \mat{0}$, $\Delta \mat{F} \mat{V}_{f'} \mat{V}_{f'}^{\rm T} = \mat{0}$, and $\mat{U}_{f'}^{\rm T} \Delta \mat{F} \mat{V}_{f'} = \mat{0}$, and thus the second bracketed terms in Eqs.~(\ref{eq:SVP-Lp})---(\ref{eq:SVP-Dp}) are equal to $\Delta \mat{F}$ due to orthogonality. 
The incomplete foreground model already discussed in \S\ref{S:rf} is such an example, where $\mat{F}' = \mat{F}_1$ and $\Delta \mat{F} = \mat{F}_2$, and $\mat{U}_{f'} = \mat{U}_{f_1}$, $\mat{U}_{\Delta f} = \mat{U}_{f_2}$. Like in \S\ref{S:rf}, the ``orthogonally'' biased foreground model just introduces an additional term $\Delta \mat{F}$ in the final results. 

More generally, the impact of biased foreground model should be between the above two extreme cases, i.e.~the foreground mixing introduced by the SVP method should not be larger than the offset $\Delta \mat{F}$ of the foreground model. For the rest of this paper, we neglect this systematic error introduced by biased foreground model.

\section{Simulation Test} \label{S:ex}

\begin{figure*}
  \centering
  \includegraphics[trim=180 10 180 10,clip,width=0.4\textwidth]{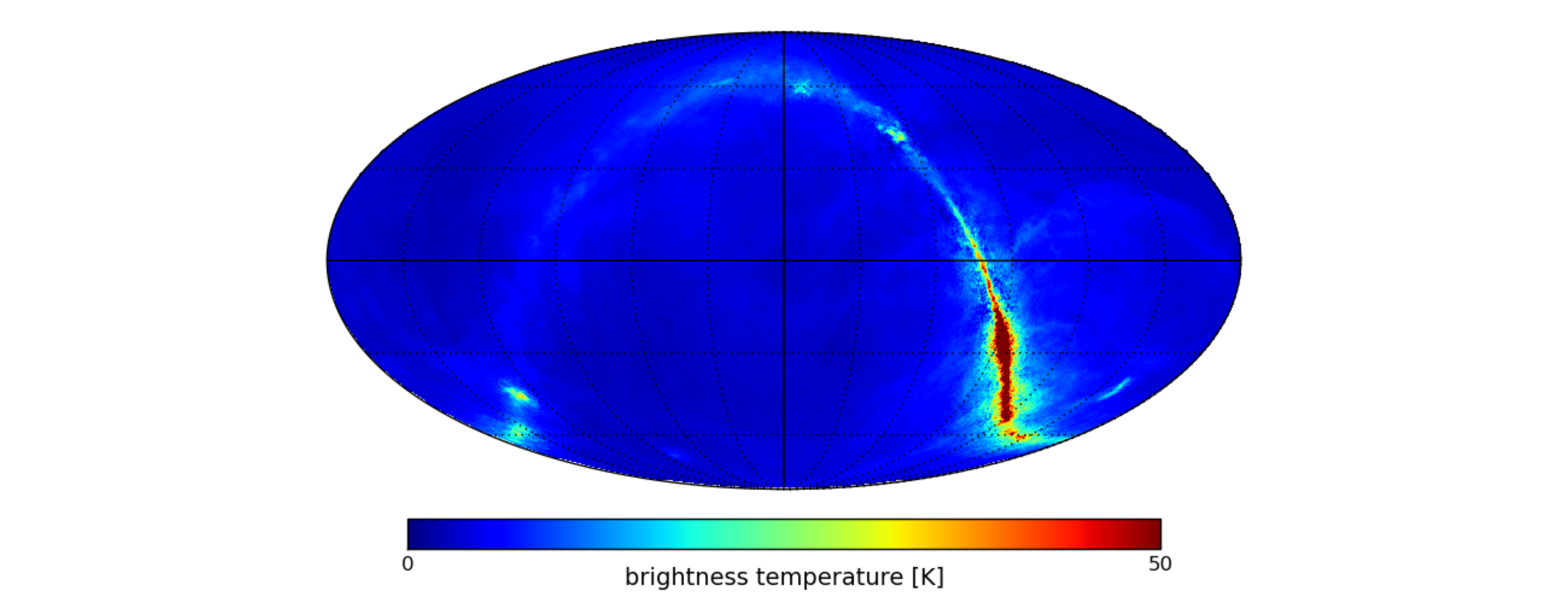}
  \includegraphics[trim=180 10 180 10,clip,width=0.4\textwidth]{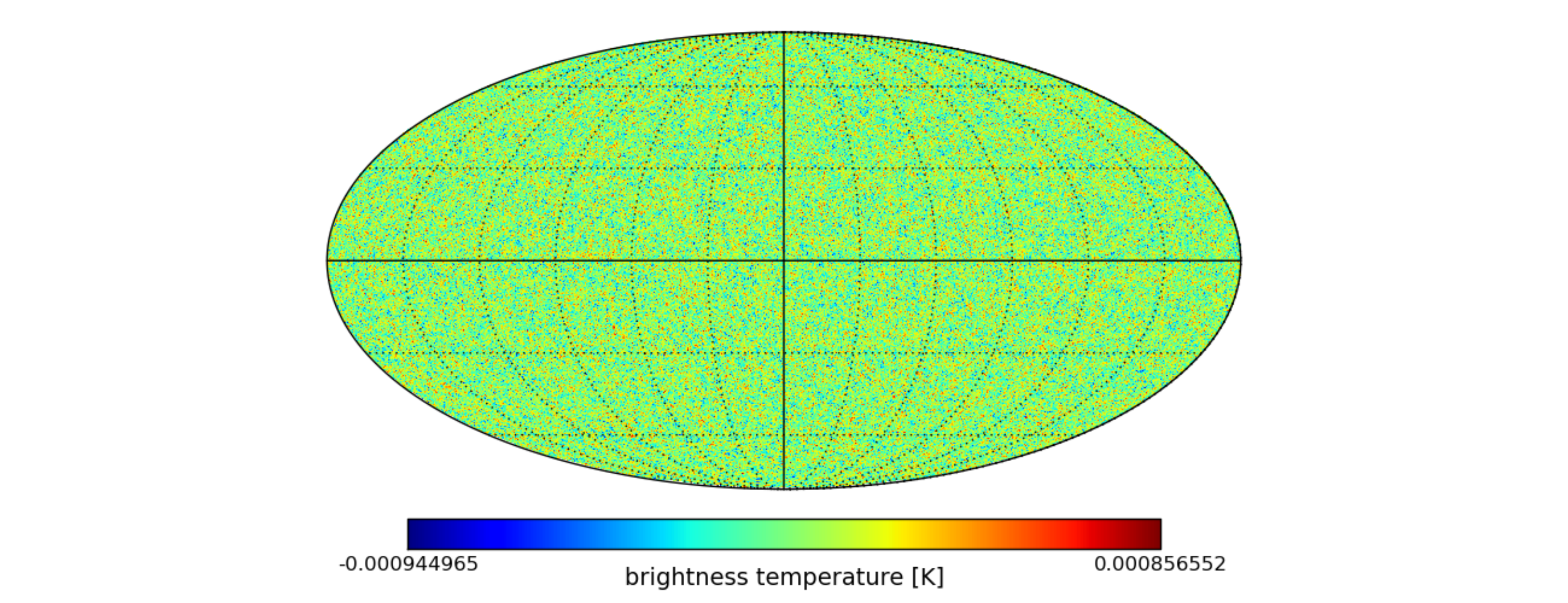} \\
  \includegraphics[trim=180 10 180 10,clip,width=0.4\textwidth]{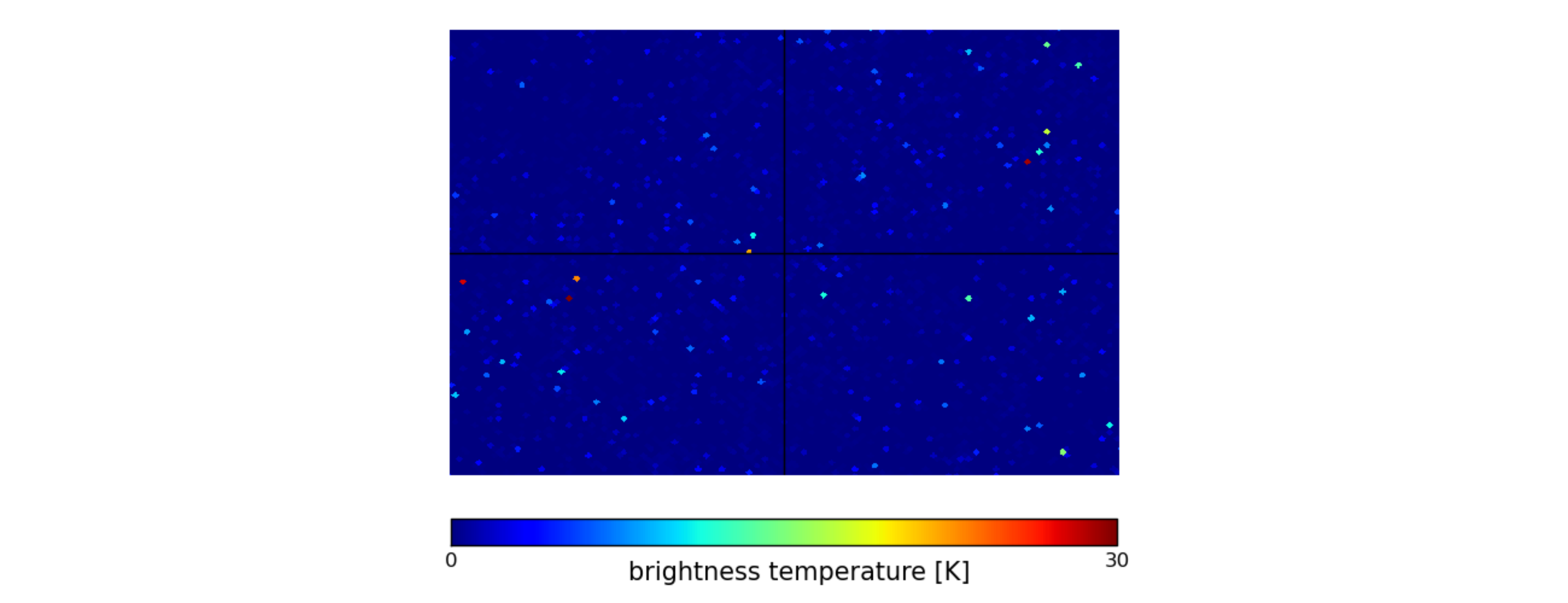}
  \includegraphics[trim=180 10 180 10,clip,width=0.4\textwidth]{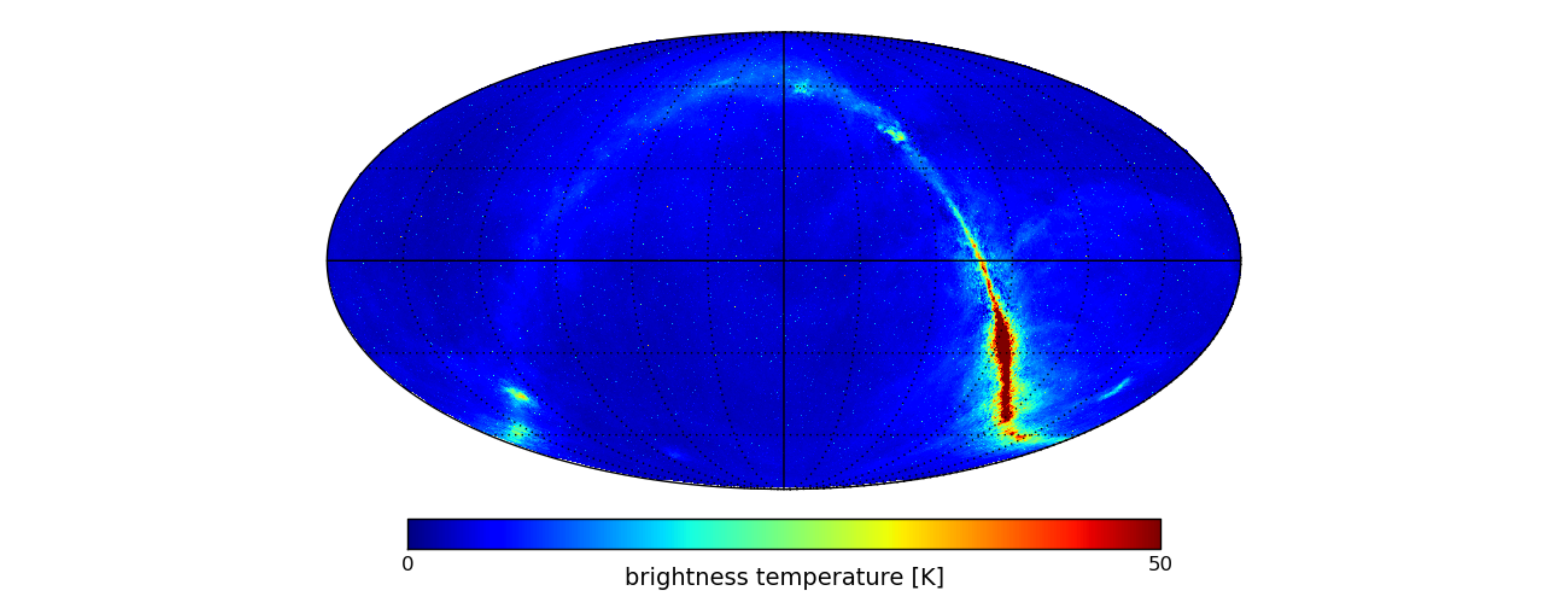}
  \caption{Input sky maps at 750 MHz. We show the brightness temperature of the simulated Galactic synchrotron emission (top left), the cosmic 21~cm signal (top right), extragalactic point sources (bottom left), and the sum of all components (bottom right), respectively. For better visualization effect, we only show a small patch of the sky, ($-15\degree \le \alpha \le 15\degree$, $-10\degree \le \delta \le 10\degree$) for the map of point sources.}
  \label{fig:maps}
\end{figure*}

\subsection{Simulation Setup} \label{S:sim}

In this section, we test the performance of SVP in terms of recovery errors with simulation data. We use the {\tt CORA}\footnote{\url{https://github.com/radiocosmology/cora}}
\citep{Shaw2014,Shaw2015} package to generate the simulated dataset, which includes the 
H{~\sc i} 21~cm signal and the mock foregrounds with two dominant components ---  
galactic synchrotron emission and extragalactic point sources. We assume the Planck 2013 cosmological model \citep{Ade2014}.

For the H{~\sc i} emission, the 21~cm power spectrum is given by
\begin{equation} \label{eq:PTb}
  P_{T_{b}}(\vec{k}; z, z') = \bar{T}_{b}(z) \bar{T}_{b}(z') (b + f
  \mu^{2})^{2} P_{m}(k; z, z'),
\end{equation}
where $b$ is the bias, $f$ is the growth rate, and $P_{m}\left(k ; z, z^{\prime}\right)=P(k) D_{+}(z) D_{+}\left(z^{\prime}\right)$ is the real-space matter power spectrum, $D_{+}$ is the growth factor normalized such that $D_{+}(0)=1$. The mean brightness
temperature takes the form \citep{Chang2008}
\begin{equation}\label{eq:Tbz}
  \bar{T}_{b}(z) = 0.3 \, \left( \frac{\Omega_{\text{H{~\sc i}}}}{
      10^{-3}} \right)   \left( \frac{1 +  z}{2.5} \right)^{1/2}    
      \left[ \frac{\Omega_{\rm m} + (1 + z)^{-3}
      \Omega_{\Lambda}}{0.29} \right]^{-1/2}\,\text{mK}.
\end{equation}
We adopt the typical values of {\tt CORA} parameters:  
$\Omega_{\text{H{~\sc i}}} b = 6.2 \times 10^{-3}$  \citep{Switzer2013} and $b = 1$.

The 21~cm angular power spectrum is given by \citep{Datta2007}
$C_{l}(\Delta \nu) \propto \int k^{2} dk j_{l}(k \chi) j_{l}(k \chi') P_{T_{b}}(\vec{k}; z, z')$, 
where $\Delta \nu = \nu' - \nu$. Here, $\chi$ ($\chi'$) is the comoving distance to redshift $z$ ($z'$) that corresponds to the frequency $\nu$ ($\nu'$).  In the flat-sky approximation that is accurate to the percentage level, the 21~cm angular power spectrum is \citep{Datta2007,Shaw2014}
\begin{equation} \label{eq:Clzz}
  C_{l}(z, z') = \frac{1}{\pi \chi \chi'} \int_{0}^{\infty}
  dk_{\parallel} \cos(k_{\parallel} \Delta \chi) P_{T_{b}}(\vec{k};
  z, z'),
\end{equation}
where $\Delta \chi = \chi-\chi'$. In the integration, the vector $\vec{k} = (k_{\parallel},k_\perp=l/\bar{\chi})$, where $\bar{\chi}=(\chi+\chi')/2$.

To model the foregrounds, for simplicity, we only consider the foregrounds with two main sources at low frequencies --- the galactic synchrotron radiation and the extragalactic radio point sources, and ignore other minor sources such as free-free emission and dust emission. 
The angular power spectra of the foregrounds with these two main sources can be modeled in the form of
\begin{equation} \label{eq:clmdl}
  C_{l}(\nu, \nu') = A \left( \frac{l}{100} \right)^{-\alpha} \left(
    \frac{\nu \nu'}{\nu_{0}^{2}} \right)^{-\beta}
  e^{-\frac{1}{2\xi_{l}^{2}}\ln^{2}(\nu / \nu')}\,,
\end{equation}
where we choose the pivot frequency $\nu_0 = 130$ MHz \citep{Centos2005}. In principle, $\xi_{l}$ is a function of $l$. For simplicity, {\tt CORA} assumes that $\xi$ is independent of $l$. 
We use the recalibrated model parameters for the 700-800 MHz band of H{~\sc i} intensity mapping experiment. The {\tt CORA} package also implemented the polarized emission model but 
for simplicity we only consider the total intensity model. 
We follow \citet{Shaw2014} for choosing the values of the model parameters, as listed in Table~\ref{tab:clp}.

To generate the galactic synchrotron emission, 
{\tt CORA} uses the processed 408 MHz Haslam 
map (with bright point sources and striping removed) \citep{Haslam1982,Remazeilles2015}
as a template, and extrapolate it to other frequencies using a spectral 
index from the Global Sky Model (GSM; \citealt{Costa2008}),  
with a Gaussian random realization that is consistent with the angular power spectra of the foregrounds (Eq.~\ref{eq:clmdl}) and adds fluctuations in frequency and on small angular scales.  
The extragalactic point sources simulations
come from three components: a population of bright point
sources ($S > 10$~Jy at 151~MHz), a synthetic population of dimmer
sources down to 0.1~Jy at 151~MHz, and an unresolved background of
dimmer sources ($S < 0.1$~Jy) modeled as a Gaussian random
realization from Eq.~(\ref{eq:clmdl}) with the point source model
parameters listed in \autoref{tab:clp}.

\begin{table}
  \centering
      \caption{Parameter Values of the Foreground Model}
  \begin{tabular}{llllll}
    \hline\hline
    Component & Polarization & $A (\text{K}^{2})$ & $\alpha$ & $\beta$ & $\xi$ \\
    \hline
    Galaxy & TT & $6.6 \times 10^{-3}$ & 2.80 &  2.8 & 4.0 \\
    Point sources & TT & $3.55 \times 10^{-4}$ & 2.10 &  1.1 & 1.0 \\
    \hline\hline
  \end{tabular}
  \label{tab:clp}
\end{table}

We generate all components of foregrounds in each of 256 frequencies uniformly sampled between 700 and 800 MHz. For visualization, we show these simulated components only at the central frequency 750 MHz in Figure~\ref{fig:maps}.

To include instrumental effects in the real observation, the generated sky maps are convolved with a symmetric circular frequency-dependent Gaussian beam with FWHM $= 1.22 \,\lambda / D$, where $\lambda$ is the observing wavelength and $D$ is the diameter of a telescope. We assume $D = 100$~m, which is about the dish size of the GBT, currently the largest fully-steerable telescope \citep{Chang2010,Masui2013,Switzer2013}, and also the optimal dish size for the mid-redshift 21~cm intensity mapping experiments \citep{Chang2008,Seo2010,Ansari2012}.

\begin{figure}
  \centering
  \includegraphics[width=0.4\textwidth]{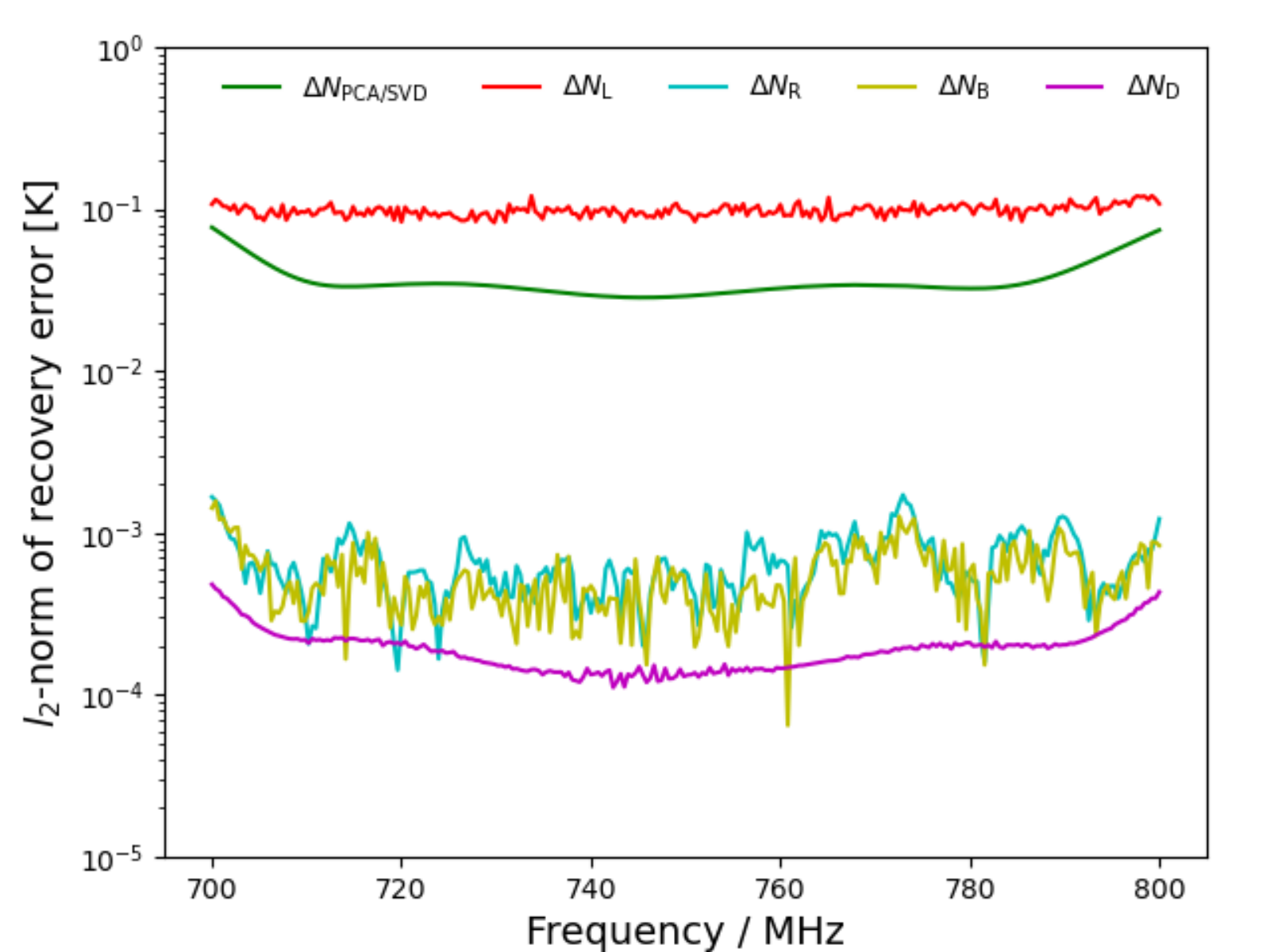}
  \caption{The $l_{2}$-norm of recovery error $\| \Delta\mat{N} \|$ as a function of frequency for the PCA/SVD estimator (green), and for the SVP estimators $\mat{N}_{\rm L}$ (red), $\mat{N}_{\rm R}$ (cyan), $\mat{N}_{\rm B}$ (yellow), and $\mat{N}_{\rm D}$ (magenta). For SVP estimators, we assume all modes of the left and/or right singular vectors of the foregrounds are known. 
  }
  \label{fig:l2}
\end{figure}


\begin{figure}
  \centering
  \includegraphics[width=0.4\textwidth]{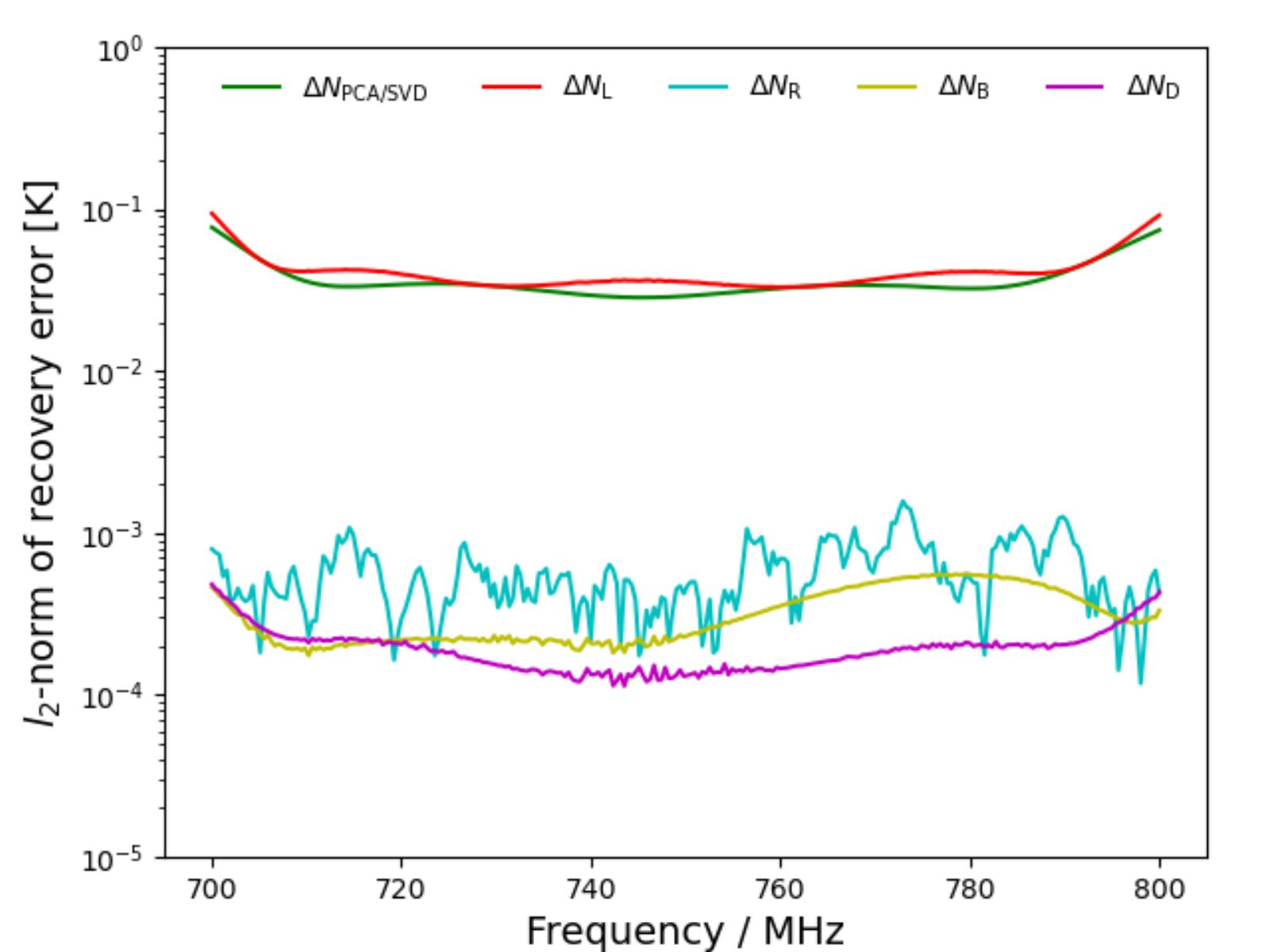}
  \caption{Same as Figure~\ref{fig:l2}, but for the SVP estimators, we only exploit the largest five left and/or right singular vectors of the foregrounds, i.e.\ $\mat{U}_{f_{1}}$ and $\mat{V}_{f_{1}}$.}
  \label{fig:l25}
\end{figure}

\begin{figure}
  \centering
  \includegraphics[width=0.4\textwidth]{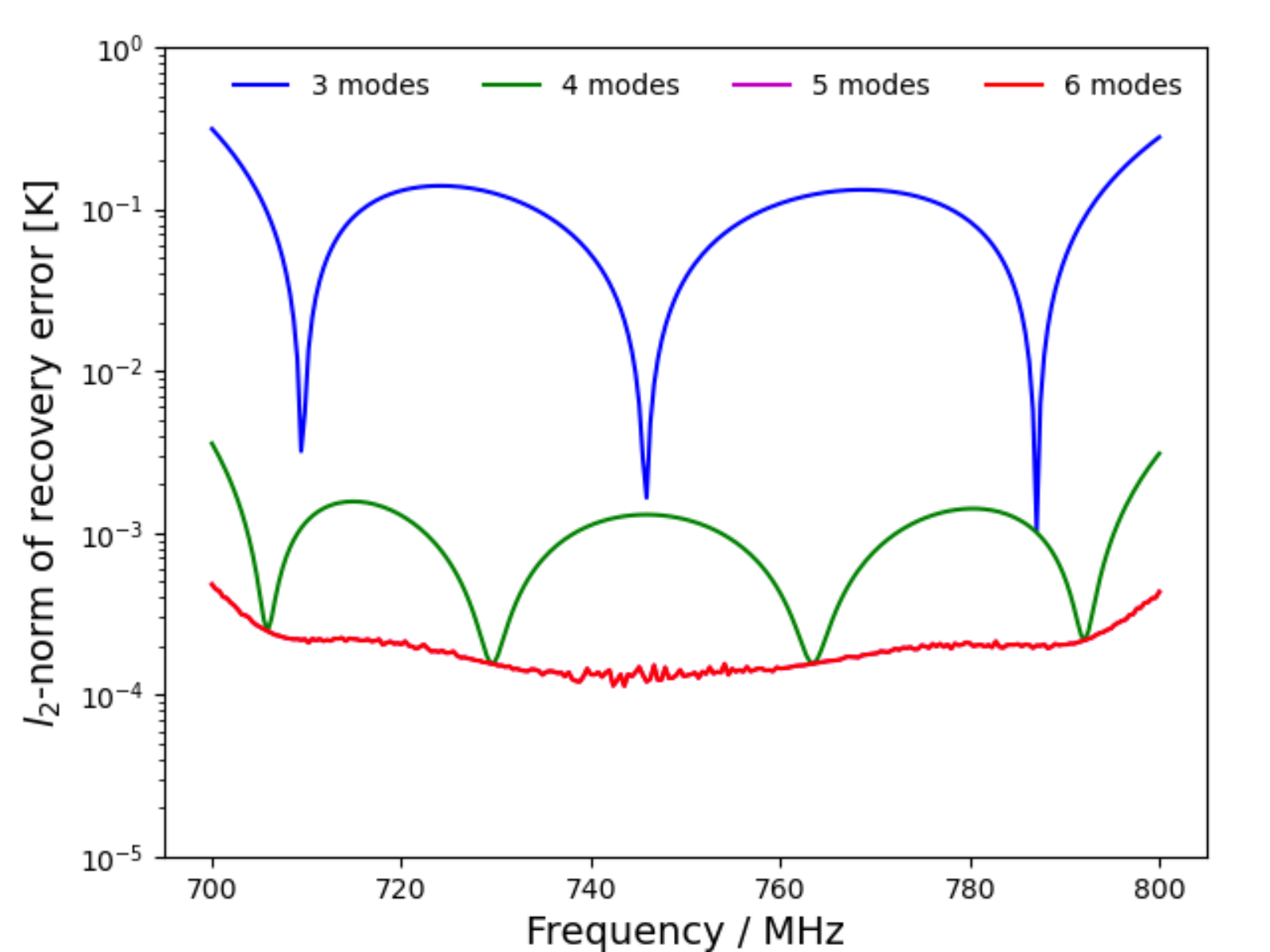}
  \caption{The $l_{2}$-norm of recovery error $\| \Delta\mat{N} \|$ as a function of frequency for the SVP estimator $\mat{N}_{\rm D}$, when we only exploit the largest three (blue), four (green), five (magenta), and six (red) left and right singular vectors of the foregrounds, respectively. Note that the magenta curve overlaps with the red one.}
  \label{fig:l2nd}
\end{figure}

\subsection{Results} \label{S:esl}

We apply the (blind) PCA/SVD method and the (semi-blind) SVP method to the simulated dataset, and test their performance using the measures of the $l_2$-norm, power spectrum, and Pearson correlation coefficient, as follows. For the PCA/SVD, the largest five PCA modes are removed, because these five modes dominate our dataset when instrumental effects are included. For the SVP, we assume that the left and/or right singular vectors ($\mat{U}_{f}$ and/or $\mat{V}_{f}$) for all or at least some largest modes are known {\it a priori}.

\subsubsection{$l_{2}$-norm}

The $l_{2}$-norm of a $1\times p$ vector $\vec{x}$ is defined as
$$ \| \vec{x} \| = \sqrt{\vec{x} \cdot \vec{x}^{T}} = \sqrt{\sum_{i=1}^{p} x_{i}^{2}}\,. $$ 
At a given frequency, the recovery error can be treated as a $1\times p$ vector, so we compute the $l_{2}$-norm of recovery error as a function of frequency in Figures~\ref{fig:l2} and \ref{fig:l25}, as a statistical measure of the performance of different estimators. A smaller $l_{2}$-norm means the better recovery of the true signal. 

In Figure~\ref{fig:l2}, we assume that the left and/or right singular vectors ($\mat{U}_{f}$ and/or $\mat{V}_{f}$) for {\it all} modes are known {\it a priori}. We further consider the scenario of incomplete information of singular vectors in Figure~\ref{fig:l25}, where we assume that only the largest five left and/or right singular vectors of the foregrounds ($\mat{U}_{f_1}$ and/or $\mat{V}_{f_1}$) are known {\it a priori}. 
Both Figures~\ref{fig:l2} and \ref{fig:l25} show that the $l_{2}$-norm of recovery error for $\mat{N}_{\rm D}$ is the smallest ($\sim 10^{-4}$) at almost all frequencies; the $l_{2}$-norm of $\mat{N}_{\rm B}$ and that of $\mat{N}_{\rm R}$ are comparable ($\sim$ a few of $ 10^{-4}$) but both larger than $\mat{N}_{\rm D}$; the $l_{2}$-norm of $\mat{N}_{\rm L}$ and that of PCA/SVD are the largest ($\sim$ a few of $ 10^{-2}$). This demonstrates that the SVP estimators except for $\mat{N}_{\text{L}}$\footnote{To understand the comparable results of the PCA/SVD and $\mat{N}_{\text{L}}$, we note that while the SVP method can reduce the foreground mixing (for incomplete information of singular vectors) even down to zero (for complete information of all modes), the signal loss in this process of foreground subtraction is not necessarily smaller than in the blind PCA/SVD. That is indeed the motivation of our tests with simulations.} perform better than the PCA/SVD estimator, generally. In particular, the $\mat{N}_{\rm D}$ estimator can reduce the $l_{2}$-norm of recovery error by two orders of magnitude over the PCA/SVD method. Also, these results agree with the relations found in Appendix~\ref{S:csl}: $\|\Delta \mat{N}_{\text{D}}\| \le \|\Delta \mat{N}_{\text{B}} \|\le \|\Delta \mat{N}_{\text{L}}\|$ and $\|\Delta \mat{N}_{\text{D}}\| \le \|\Delta \mat{N}_{\text{B}}\| \le \|\Delta \mat{N}_{\text{R}}\|$. Roughly speaking, this implies that exploiting additional information in the singular vectors can improve the accuracy of signal recovery over the blind PCA/SVD method for foreground subtraction, and exploiting more information (in both left and right singular vectors) is better than only using partial information (in either left or right singular vector).  

We note that if only a small number of modes of singular vectors corresponding to the largest singular values are exploited in SVP, there is residual foreground mixing (the $\mat{F}_2$ term in Eq.~\ref{eq:svpinc}) that contributes to the recovery error. However, comparing the results of Figures~\ref{fig:l2} and \ref{fig:l25}, we find the $l_2$-norm of recovery error for the SVP estimators with only the largest five modes of singular vectors is at the same level as that with full information of singular vectors. This is encouraging because in real observations, it is more likely to only obtain the information of the largest few modes than that of all modes. 

We further explore the optimal number of retained modes for the SVP estimator $\mat{N}_{\rm D}$ in Figure~\ref{fig:l2nd}, and find that the $l_{2}$-norm of recovery error converges when the number of largest retained modes exceeds five. This convergence allows realistic application of the SVP method.


\begin{figure}
  \centering
  \includegraphics[width=0.4\textwidth]{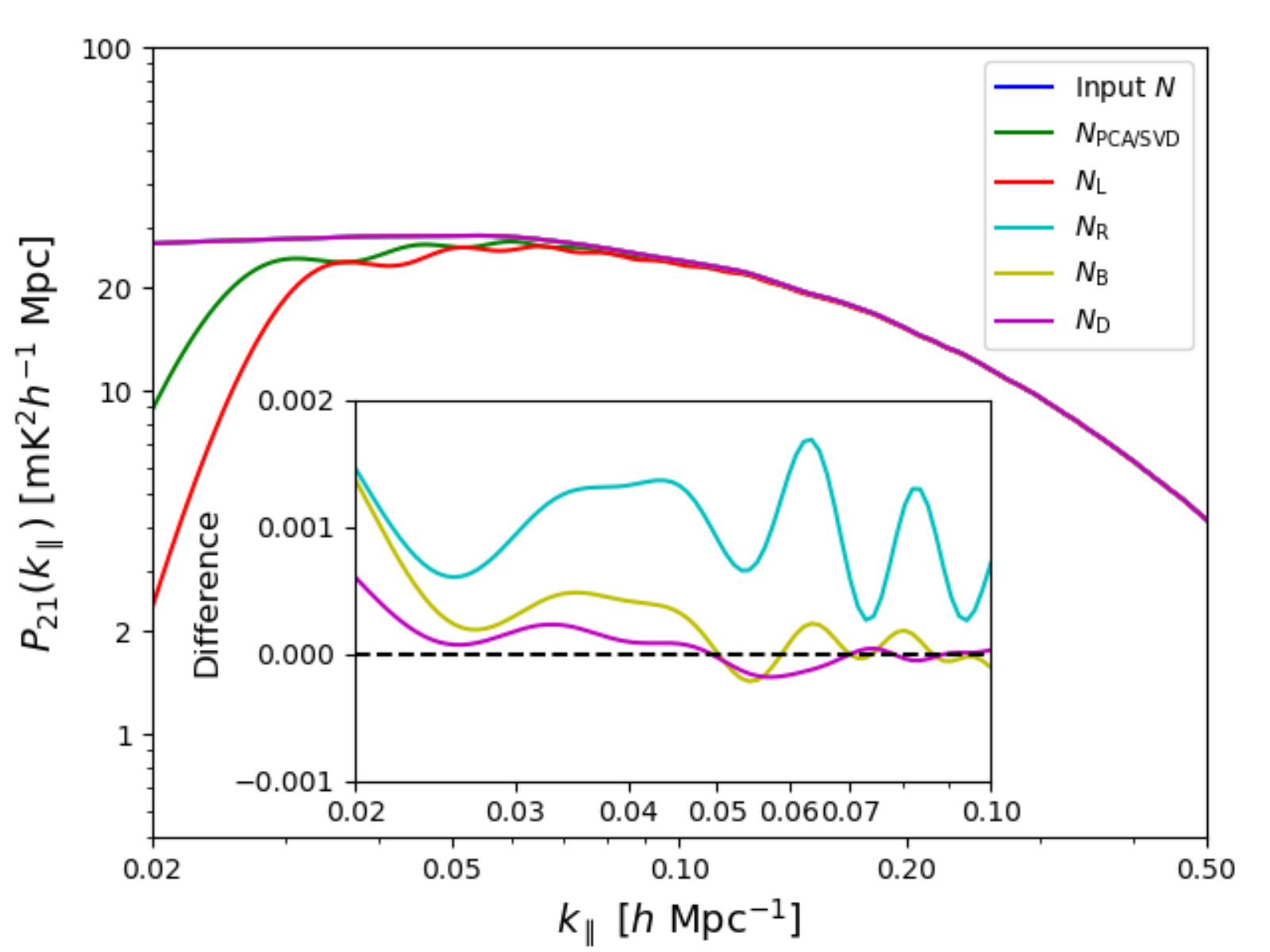}
  \caption{The 1D power spectrum along the line-of-sight. We show the result of the input, true signal $\mat{N}$ (blue), and the recovered signal using the PCA/SVD estimator $\mat{N}_{\rm PCA/SVD}$ (green) and the SVP estimators, $\mat{N}_{\rm L}$ (red), $\mat{N}_{\rm R}$ (cyan), $\mat{N}_{\rm B}$ (yellow), and $\mat{N}_{\rm D}$ (magenta), respectively. For the SVP estimators, we only exploit the largest five left and/or right singular vectors of the foregrounds, i.e.\ $\mat{U}_{f_{1}}$ and $\mat{V}_{f_{1}}$.  
  The cyan, yellow and magenta curves overlap with the blue curve, and we show their difference with respect to the input signal in inset.}
  \label{fig:pk}
\end{figure}

\begin{figure}
  \centering
  \includegraphics[width=0.45\textwidth]{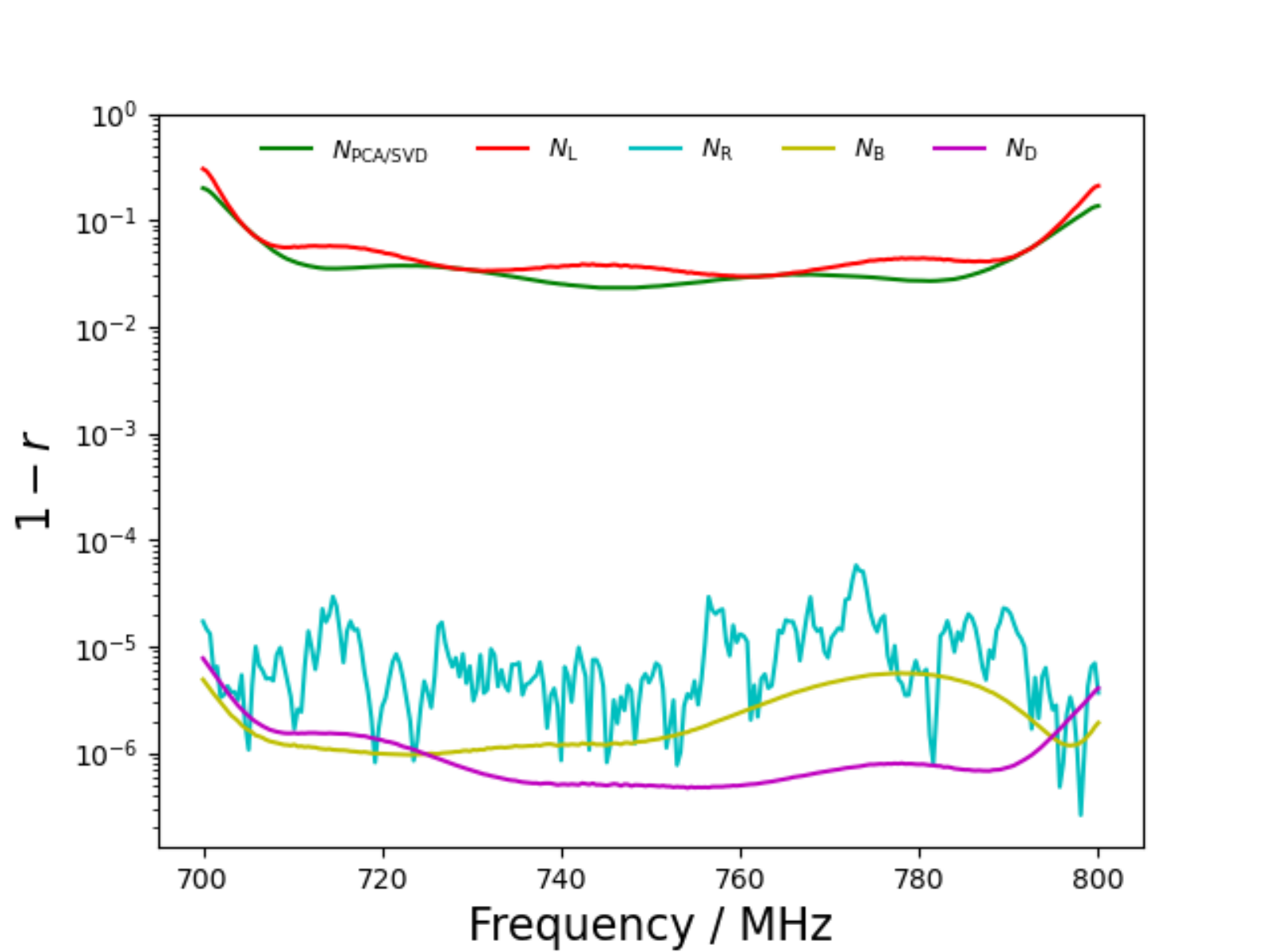}
  \caption{The Pearson correlation coefficient $r$ between the input, true signal $\mat{N}$ and the recovered signal using the PCA/SVD estimator $\mat{N}_{\rm PCA/SVD}$ (green) and the SVP estimators, $\mat{N}_{\rm L}$ (red), $\mat{N}_{\rm R}$ (cyan), $\mat{N}_{\rm B}$ (yellow), and $\mat{N}_{\rm D}$ (magenta), respectively. We plot $1-r$ as a function of frequency for visualization purpose. For the SVP estimators, we only exploit the largest five left and/or right singular vectors of the foregrounds, i.e.\ $\mat{U}_{f_{1}}$ and $\mat{V}_{f_{1}}$. }
  \label{fig:r}
\end{figure}

\subsubsection{Power Spectrum}

In Figure~\ref{fig:pk}, we plot the 1D power spectrum along the line-of-sight (LoS) of the 21~cm signal, $P_{21}(k_{\parallel})$, for the input, true signal $\mat{N}$ as the benchmark, and for the recovered signal using the PCA/SVD and SVP estimators. For the SVP estimators, we only exploit the largest five left and/or right singular vectors of the foregrounds, i.e.\ $\mat{U}_{f_1}$ and/or $\mat{V}_{f_1}$. We find that these estimators can recover the input power spectrum well at small scales $k_\parallel > 0.1\,h\,{\rm Mpc}^{-1}$. On the other hand, at large scales, while the PCA/SVD method and the $\mat{N}_{\rm L}$ estimator lose the power significantly, the $\mat{N}_{\rm D}$, $\mat{N}_{\rm B}$ and $\mat{N}_{\rm R}$ estimators can recover the input power spectrum at high accuracy. In comparison, the $\mat{N}_{\rm D}$ estimator performs the best, with the absolute difference $< 0.001\,{\rm mK}^2\,h^{-1}\,{\rm Mpc}$ and relative error $< 0.01\%$. These results are consistent with the findings in the $l_2$-norm test.


\subsubsection{Pearson Correlation Coefficient}

The Pearson correlation coefficient $r$ is a statistical measure of the degree of linear correlation between two signals. For the input $1\times p$ vector $\vec{x}$ and the recovered vector $\hat{\vec{x}}$, it is defined as 
\begin{equation}
    r = \frac{\Delta\vec{x} \cdot \Delta\hat{\vec{x}}^{T}}{\sqrt{\Delta\vec{x} \cdot \Delta\vec{x}^{T}} \, \sqrt{\Delta\hat{\vec{x}} \cdot \Delta\hat{\vec{x}}^{T}} }
    = \frac{\sum\nolimits_{i}(x_{i} -
  \bar{x})(\hat{x}_{i} -
  \bar{\hat{x}})}{\sqrt{\sum\nolimits_{i}(x_{i}
    - \bar{x})^{2}} \, \sqrt{\sum\nolimits_{i}(\hat{x}_{i}
    - \bar{\hat{x}})^{2}}}    \,,
\end{equation}
where $\Delta\vec{x} = \vec{x} - \bar{x}$, and $\Delta\hat{\vec{x}} = \hat{\vec{x}} 
 - \bar{\hat{x}}$. Here, $\bar{x}$ is the mean of $\vec{x}$, and $\bar{\hat{x}}$ is the mean of $\hat{\vec{x}} $. The value of $r$ is close to unity if two signals are very correlated. 

We compute the Pearson correlation coefficient $r$ between the input, true 21~cm signal and the recovered signal using different estimators. For visualization purpose, we plot the value of $1-r$ in Figure~\ref{fig:r}, because the values of $r$ are all close to unity, but how $1-r$ is close to zero shows the degree of similarities between the recovered signal and the true signal. We find that the values of $1-r$ for the $\mat{N}_{\rm D}$, $\mat{N}_{\rm B}$ and $\mat{N}_{\rm R}$ estimators are comparable ($\sim 10^{-5} - 10^{-6}$) and significantly smaller than that of the PCA/SVD and the $\mat{N}_{\rm L}$ estimator ($\sim 10^{-1} - 10^{-2}$). Here we only exploit the largest five left and/or right singular vectors of the foregrounds for the SVP estimators. These results are consistent with the results in the $l_2$-norm and power spectrum tests.

\section{Conclusions} \label{S:con}
Foreground subtraction is one of the key challenges to the 21~cm observations due to the fact that the foregrounds are four to five orders of magnitudes larger than cosmic 21~cm signal. In this paper, we show that the PCA method and the SVD method, which are widely employed for foreground subtraction, are actually equivalent in principle. We also provide the conditions in which the PCA/SVD method can separate the foregrounds from the 21~cm signal completely, i.e.\ with zero residual foreground mixing and zero signal loss. Nevertheless, we point out that in general the foreground mixing and signal loss are unavoidable for the PCA/SVD method, because those conditions are hardly satisfied in practice. 

In this paper, we propose a new class of {\it semi-blind} method for foreground subtraction, based on the PCA/SVD method, called the {\it Singular Vector Projection}. The SVP method is semi-blind in the sense that it exploits {\it a priori} information of the left and/or right singular vectors of the foregrounds --- if only the left (right) singular vector is known, then the estimator $\mat{N}_{\rm L}$ ($\mat{N}_{\rm R}$) can be employed; if both left and right singular vectors are known, then two estimators, $\mat{N}_{\rm B}$ and $\mat{N}_{\rm D}$, can be employed. The virtue of SVP is that, in principle, the residual foreground mixing is zero for all four SVP estimators, if the information of singular vectors for {\it all} modes are known. 

We generate the mock maps of the 21~cm signal and the foregrounds from simulations, and use them to test the performance of the SVP estimators and the standard PCA/SVD estimator in terms of the $l_2$-norm of recovery error, the LoS power spectrum of the 21~cm signal, and the Pearson correlation coefficient between the input, true signal and the recovered signal. We find that while the results of $\mat{N}_{\rm L}$ estimator are comparable to those of the PCA/SVD method, the other SVP estimators ($\mat{N}_{\rm R}$, $\mat{N}_{\rm B}$ and $\mat{N}_{\rm D}$) can significantly improve the accuracy of recovery by orders of magnitude. In particular, the $\mat{N}_{\rm D}$ estimator performs the best in general. 

We also consider the more realistic scenario of incomplete foreground information, in which only the largest few modes of the left and/or right singular vectors of the foregrounds are known {\it a priori}. In this case, there is residual foreground mixing due to the other small modes of residual foregrounds. However, the accuracy of recovery with the SVP estimators, if the largest five modes of the foregrounds are given in our demonstration, is not degraded with respect to the case with the information of all-mode singular vectors. This indicates that the SVP estimator reaches a balance between signal loss and residual foreground mixing. 

Regarding the availability of {\it a priori} information, the left and right singular vectors can be solved by eigen-decomposition of the frequency and pixel covariance matrices of the foregrounds, respectively. Specifically, accurate frequency covariance matrix of the foregrounds can be obtained with low cost by observations at other frequency bands. Our test results reflect the fact that the frequency spectrum information of the foregrounds is insufficient for the 21~cm experiments to reach high precision, and the spatial information of the foregrounds should be taken into account as well. 

However, accurate foreground modeling is still a challenge to 21~cm observations. For example, Haslam and other maps may have offsets and/or incorrect scaling and may typically double-count unresolved or point-like sources \citep{Monsalve2021}. With the SVP, the impact of biased foreground model is restricted to be smaller than the offset between the foreground model and the true foreground. 
In particular,
the SVP estimators are independent of the overall magnitude of the covariance matrix, which can be highly biased. This may be an advantage against some other semi-blind methods that depend on the overall magnitude of the foreground covariance matrix, e.g.\ the KL transform method \citep{Shaw2014,Shaw2015}.

When such {\it a priori} information of the foregrounds is available, our paper suggests that the SVP estimators can improve the recovery results significantly over the standard PCA/SVD method which is blind against the prior information. In particular, the right singular vector of the foregrounds can help improve the foreground subtraction in a more effective manner than the left singular vector. Furthermore, combining both left and right singular vectors, the $\mat{N}_{\rm D}$ estimator works the best. These SVP estimators provide a new, effective approach for 21~cm observations to remove foregrounds and uncover the physics in the cosmic 21~cm signal.


\section*{Acknowledgements}
This work is supported by National SKA Program of China (grant No.~2020SKA0110401), NSFC (grant No.~11821303, 11633004), National Key R\&D Program of China (grant No.~2018YFA0404502), the MOST inter-government cooperation program China-South Africa Cooperation Flagship project (grant No.~2018YFE0120800), the Chinese Academy of Sciences (CAS) Frontier Science Key Project (grant No.~QYZDJ-SSW-SLH017), the CAS Strategic Priority Research Program (grant No.~XDA15020200). We thank Fengquan Wu and Yichao Li for useful discussions and helps. We acknowledge the Tsinghua Astrophysics High-Performance Computing platform at Tsinghua University for providing computational and data storage resources that have contributed to the research results reported within this paper. 

\software{CORA \citep{Shaw2014}, PyGSM \citep{Costa2008,Zheng2017}, h5py
\citep{Collette2021}, healpy \citep{Zonca2019}, Matplotlib
\citep{Hunter2007}, NumPy \citep{Harris2020}, SciPy
\citep{Virtanen2020}}

\appendix

\section{Inequalities for Signal Loss} \label{S:csl}

In this section, we prove some inequalities for the signal loss using the SVP estimators. These inequalities focus on the comparison of the ``magnitude'' of signal loss. 

We quantify the ``magnitude'' of signal loss using the Frobenius norm \citep{Noble1977}. For a $m \times n$ matrix $\mat{A}$, its Frobenius norm $\| \mat{A} \|_{\rm F}$ is defined as
\begin{equation} \label{eq:AF}
  \| \mat{A} \|_{\rm F} = \sqrt{\text{Tr}(\mat{A}^{\rm T} \mat{A})} = \sqrt{\sum_{i=1}^m \sum_{j=1}^n |a_{ij}|^2}.
\end{equation}

We first prove three inequalities of Frobenius norm. 

(1)
\begin{equation} \label{eq:ABF}
  \| \mat{A} \mat{B} \|^2_{\rm F} \le \| \mat{A} \|^2_{\rm F} \, \| \mat{B} \|^2_{\rm F}.
\end{equation}
The proof is as follows.
\begin{align}
  \| \mat{A} \mat{B} \|^2_{\rm F} &= \sum_{i=1}^m \sum_{j=1}^p \left|\sum_{k=1}^n a_{ik} b_{kj} \right|^2 \notag \\
    &\le \sum_{i=1}^m \sum_{j=1}^p \left[ \left(\sum_{k=1}^n |a_{ik}|^2 \right) \left(\sum_{l=1}^n |b_{lj}|^2 \right) \right]  \notag \\
    &= \left(\sum_{i=1}^m \sum_{k=1}^n |a_{ik}|^2 \right) \left(\sum_{l=1}^n \sum_{j=1}^p |b_{lj}|^2 \right)\notag \\
    &= \| \mat{A} \|^2_{\rm F} \, \| \mat{B} \|^2_{\rm F}. \label{eq:ABFp}
\end{align}
The second line in Eq.~(\ref{eq:ABFp}) uses the Cauchy-Schwarz inequality. 

(2) For a square matrix $\mat{S}$, it is straightforward by its definition to prove 
\begin{equation}
\| \mat{S}_{\text{diag}} \|_{\rm F} \le \| \mat{S} \|_{\rm F}\,.    
\end{equation}

(3) For a partial orthogonal matrix $\mat{U}$, 
\begin{equation}
  \| \mat{A} \mat{U} \|_{\rm F} \le \| \mat{A} \|_{\rm F}\,.  
\end{equation}
The proof is as follows.
\begin{align}
  \| \mat{A} \mat{U} \|^2_{\rm F} &= \text{Tr}(\mat{U}^{\rm T} \mat{A}^{\rm T} \mat{A} \mat{U}) 
    = \text{Tr}(\mat{A}^{\rm T} \mat{A} \mat{U} \mat{U}^{\rm T}) \notag \\
    &= \sum_i \sum_j (\mat{A}^{\rm T} \mat{A})_{ij} (\mat{U} \mat{U}^{\rm T})_{ji} \notag \\
    &\le \sum_i \sum_j (\mat{A}^{\rm T} \mat{A})_{ij} \delta_{ji} 
    = \sum_i (\mat{A}^{\rm T} \mat{A})_{ii} \notag \\
    &= \text{Tr}(\mat{A}^{\rm T} \mat{A}) = \| \mat{A} \|^2_{\rm F}\,. \label{eq:AUF}
\end{align}
Using these properties, we have
\begin{align}
 \| \mat{N}_{\rm B}^{\text{loss}} \|^2_{\rm F} &= \| \mat{U}_{f} \mat{U}_{f}^{\rm T} \mat{N} \mat{V}_{f} \mat{V}_{f}^{\rm T} \|^2_{\rm F} \notag \\
  &= \text{Tr}(\mat{V}_f \mat{V}^{\rm T}_f \mat{N}^{\rm T} \mat{U}_f \mat{U}^{\rm T}_{f} \mat{U}_{f} \mat{U}_{f}^{\rm T} \mat{N} \mat{V}_{f} \mat{V}_{f}^{\rm T}) \notag \\
  &= \text{Tr}(\mat{V}^{\rm T}_f \mat{N}^{\rm T} \mat{U}_{f} \mat{U}_{f}^{\rm T} \mat{N} \mat{V}_{f}) \notag \\
  &= \| \mat{U}_{f}^{\rm T} \mat{N} \mat{V}_{f} \|^2_{\rm F} \\
 &\le \| \mat{U}_{f}^{\rm T} \mat{N} \|^2_{\rm F} = \| \mat{N}_{\rm L}^{\text{loss}} \|^2_{\rm F} \,.
 \end{align}
 Similarly, 
 \begin{equation}
    \| \mat{N}_{\rm B}^{\text{loss}} \|^2_{\rm F} = \| \mat{U}_{f}^{\rm T} \mat{N} \mat{V}_{f} \|^2_{\rm F} \le \| \mat{N} \mat{V}_{f} \|^2_{\rm F} = \| \mat{N}^{\text{loss}}_{\rm R} \|^2_{\rm F}\,.
 \end{equation}
 Also, 
\begin{align}
 \| \mat{N}_{\rm D}^{\text{loss}} \|^2_{\rm F} &= \| \mat{U}_{f} (\mat{U}_{f}^{\rm T} \mat{N} \mat{V}_{f})_{\text{diag}} \mat{V}_{f}^{\rm T} \|^2_{\rm F} \notag\\
    &= \text{Tr}\left(\mat{V}_f (\mat{U}_{f}^{\rm T} \mat{N} \mat{V}_{f})^{\rm T}_{\text{diag}} \mat{U}^{\rm T}_{f} \mat{U}_{f} (\mat{U}_{f}^{\rm T} \mat{N} \mat{V}_{f})_{\text{diag}} \mat{V}_{f}^{\rm T}\right) \notag\\
    &= \text{Tr}((\mat{U}_{f}^{\rm T} \mat{N} \mat{V}_{f})^{\rm T}_{\text{diag}} (\mat{U}_{f}^{\rm T} \mat{N} \mat{V}_{f})_{\text{diag}} ) \notag\\
     &= \| (\mat{U}_{f}^{T} \mat{N} \mat{V}_{f})_{\text{diag}} \|^2_{\rm F} \notag\\
     &\le \| \mat{U}_{f}^{\rm T} \mat{N} \mat{V}_{f} \|^2_{\rm F} = \| \mat{N}^{\text{loss}}_{\rm B} \|^2_{\rm F}\,.
\end{align}

Altogether, the following inequalities are proven:  
\begin{align}
   \| \mat{N}_{\rm D}^{\text{loss}} \|_{\rm F} & \le \| \mat{N}^{\text{loss}}_{\rm B} \|_{\rm F} \le \| \mat{N}_{\rm L}^{\text{loss}} \|_{\rm F}, \label{eq:N431} \\
  \| \mat{N}_{\rm D}^{\text{loss}} \|_{\rm F} & \le \| \mat{N}^{\text{loss}}_{\rm B} \|_{\rm F} \le \| \mat{N}^{\text{loss}}_{\rm R} \|_{\rm F}. \label{eq:N432}   
\end{align}
Since there is no foreground mixing for the SVP estimators if all modes of the left and/or right singular vectors of foregrounds are known {\it a priori}, the inequalities for signal loss are equivalent to the inequalities for recovery error, 
$ \| \Delta \mat{N}_{\text{D}} \|_{\rm F} \le \| \Delta \mat{N}_{\text{B}} \|_{\rm F} \le \| \Delta \mat{N}_{\text{L}}\|_{\rm F} $ and $\| \Delta \mat{N}_{\text{D}} \|_{\rm F} \le \| \Delta \mat{N}_{\text{B}} \|_{\rm F} \le \| \Delta \mat{N}_{\text{R}}\|_{\rm F}$.

\bibliographystyle{aasjournal}
\bibliography{pca}

\end{document}